\documentclass{article}

\usepackage[nonatbib, final]{neurips_2021}

\usepackage[utf8]{inputenc} 
\usepackage[T1]{fontenc}    
\usepackage{hyperref}       
\usepackage{url}            
\usepackage{booktabs}       
\usepackage{amsfonts}       
\usepackage{nicefrac}       
\usepackage{microtype}      
\usepackage{xcolor}         
\usepackage{graphicx}
\usepackage{tikz}           
\usepackage{subfigure}
\usepackage{xr-hyper}
\usepackage{chngcntr}
\usepackage{floatrow}       
\usepackage{hyperref}

\usepackage{float}
\usepackage{wrapfig}
\restylefloat{figure}

\title{Neural Population Geometry Reveals the Role of Stochasticity in Robust Perception}

\author{Joel Dapello$^{*,1,2,3}$, Jenelle Feather$^{*,1,2,4}$, Hang Le$^{*,1}$, Tiago Marques$^{1,2,4}$\\
        \textbf{David D. Cox$^{5}$, Josh H. McDermott$^{1,2,4,6}$, James J. DiCarlo$^{1,2,4}$, SueYeon Chung$^{1,7,8}$} \\
        \\
        $^*$Equal contribution \\
        $^1$Department of Brain and Cognitive Sciences, Massachusetts Institute of Technology \\
        $^2$McGovern Institute for Brain Research, Massachusetts Institute of Technology \\
        $^3$School of Engineering and Applied Sciences, Harvard University \\
        $^4$Center for Brains, Minds and Machines, Massachusetts Institute of Technology \\
        $^5$MIT-IBM Watson AI Lab\\
        $^6$Speech and Hearing Bioscience and Technology, Harvard University\\
        $^7$Center for Theoretical Neuroscience, Columbia University\\
        $^8$Zuckerman Institute, Columbia Univeristy\\
        \\
}

\begin{document}

\maketitle

\begin{abstract}
Adversarial examples are often cited by neuroscientists and machine learning researchers as an example of how computational models diverge from biological sensory systems. Recent work has proposed adding biologically-inspired components to visual neural networks as a way to improve their adversarial robustness. One surprisingly effective component for reducing adversarial vulnerability is response stochasticity, like that exhibited by biological neurons. Here, using recently developed geometrical techniques from computational neuroscience, we investigate how adversarial perturbations influence the internal representations of standard, adversarially trained, and biologically-inspired stochastic networks. We find distinct geometric signatures for each type of network, revealing different mechanisms for achieving robust representations. Next, we generalize these results to the auditory domain, showing that neural stochasticity also makes auditory models more robust to adversarial perturbations. Geometric analysis of the stochastic networks reveals overlap between representations of clean and adversarially perturbed stimuli, and quantitatively demonstrates that competing geometric effects of stochasticity mediate a tradeoff between adversarial and clean performance. Our results shed light on the strategies of robust perception utilized by adversarially trained and stochastic networks, and help explain how stochasticity may be beneficial to machine and biological computation.\footnote{See https://github.com/chung-neuroai-lab/adversarial-manifolds for accompanying code.}
\end{abstract}

\section{Introduction}\label{submission}
In recent years, artificial neural networks (ANNs) have come to dominate both visual object recognition and auditory recognition tasks \cite{Krizhevsky2012alexnet, He2015resnet, hannun2014deep}, establishing them as leading candidate models for several domains of human perception \cite{kriegeskorte2015deep, yamins2016using, kell2018task, rajalingham2018large}. However, they still exhibit many non-human-like traits \cite{geirhos2018imagenet, feather2019metamers}. One such failure is in the existence of adversarial perturbations -- small changes to stimuli explicitly crafted to fool a model that remain imperceptible to humans \cite{Szegedy2013intriguing, Carlini2016evaluating, carlini2018audio} -- which demonstrate the fragility of some ANNs as models of biological perception.

Recently, Dapello, Marques et al. discovered that one such method, known as adversarial training \cite{Madry2017advtrain}, not only reduces the network's adversarial vulnerability but also yields network representations that are more similar to those in the primate primary visual cortex \cite{dapello2020simulating}. Motivated by this result, the authors developed VOneNets, a class of networks that simulate the primate primary visual cortex at the front of a convolutional neural network, and show improved robustness to adversarial attacks with no adversarial training. However, a number of questions remain unanswered -- in particular, while both adversarially trained networks and VOneNets have improved adversarial robustness and also greater similarity to the primate primary visual cortex, it is unclear how VOneNets achieve robustness, and in particular if the mechanism of robustness is similar to that induced by adversarial training.

A key component of robustness in VOneNets is the inclusion of stochastic representations during both training and inference, a feature inspired by biological sensory neurons which exhibit trial-to-trial variability across presentations of the same stimulus \cite{softky1993highly}. The implications of this stochasticity for information processing are open questions in neuroscience \cite{faisal2008noise, goris2014partitioning, echeveste2020cortical, davis2020spontaneous}. Pinpointing how this representational stochasticity contributes to robustness in VOneNets could drive further developments in the mechanisms of robust perception.

Here we use recently developed manifold analysis techniques from computational neuroscience \cite{chung2018classification} to look beyond accuracy and investigate the internal neural population geometry \cite{chung2021neural} of standard, adversarially trained, and biologically-inspired stochastic networks in response to clean and adversarially perturbed examples in both visual and auditory domains. We present several key findings:

\begin{itemize}
\item Using manifold analysis, we demonstrate that standard, adversarially trained, and stochastic networks each have distinct geometric signatures in response to clean and adversarially perturbed stimuli, shedding light on varied robustness mechanisms.
\item We demonstrate the generality of our findings by translating the results to a novel biologically-inspired auditory ANN, StochCochResNet50, that includes stochastic responses. Stochasticity makes auditory networks more robust to adversarial perturbations, and the underlying neural population geometry is largely consistent with that in vision networks.
\item Analysis of stochastic networks reveals a protective overlap between the representations of adversarial examples and clean stimuli. We quantitatively survey the stochasticity conditions leading to the overlap, and map a competing geometric effect mediating a trade-off between clean and adversarial performance.
\end{itemize}

\section{Related work}\label{related}

Previous work in machine learning has reported that additive noise can improve the adversarial robustness of a model: Liu et al. \cite{Liu2017random} used random noise applied to pixels and intermediate layers to improve adversarial robustness, and Cohen et al. \cite{Cohen2019smoothing} demonstrated a method to transform a model with Gaussian noise in the pixel space to one with certified robustness to attacks of a given strength. Unlike VOneNets, both \cite{Liu2017random} and \cite{Cohen2019smoothing} rely on using multiple inference passes with different noise samples to improve robustness. Further, while both of these works suggest new defenses, none analyze the properties of stochastic representations that lead to robustness.

The need to understand internal mechanisms of biological and artificial neural networks gave rise to the field of neural population geometry \cite{Kriegeskorte2021, chung2021neural}, an line of work exploring geometric properties of high-dimensional neural representations. To capture the complexity of neural representations in ANNs, various geometric analyses have been proposed, including representation similarity analysis \cite{kriegeskorte2013representational}, geodesics \cite{henaff2015geodesics}, curvature \cite{henaff2019perceptual}, and intrinsic dimensionality \cite{ansuini2019intrinsic}. Another popular approach is through supervised linear probes \cite{alain2016understanding}, i.e., linear classifiers trained on top of these representations. However, recent work has discussed how such analyses with linear classifiers are limited \cite{hewitt2019designing}, and that more structural studies are needed for investigating the internal layers of networks \cite{ivanova2021probing}. A recent theoretical development based on replica theory in statistical physics \cite{chung2016linear, chung2018classification, cohen2020separability} provides a solution to this, by formally connecting the geometry and linearly decodable information embedded in neural populations. By using this approach, our study situates the analysis of robustness mechanisms into the developing field of neural population geometry \cite{Kriegeskorte2021, chung2021neural}.

\section{Methods and experimental setup}
\label{methods}

\subsection{Replica-based manifold analysis}
In this paper, we use replica mean-field theoretic manifold analysis (MFTMA, hereafter) \cite{chung2018classification, stephenson2019untangling, cohen2020separability}, which formally connects the linear decodability of object manifolds, defined as a set of stimulus-evoked representations in neural state space \cite{dicarlo2007untangling}, to their geometrical properties. Here we provide a brief description of the key quantities (See SM 1 and \cite{chung2018classification} for a complete treatment of MFTMA). Using this framework, we analyze the size, shape, and distribution of object manifolds as they are transformed throughout the network, gaining insight into the neural population geometry underlying performance of the classification task. Specifically, given $P$ object manifolds (measured as feature point clouds) in $N$ dimensions, MFTMA returns capacity (a measure of linear separability) and the manifold dimensions, radii, and center correlations associated with capacity estimation.

\textbf{Manifold capacity} ($\alpha = P/N$) refers to the maximum number of object manifolds ($P_{max}$) that can be linearly separated given $N$ features, and characterizes the linearly-decodable object information per feature dimension. MFTMA estimates $\alpha$ through measures of the manifold dimension ($D_M$), manifold radius ($R_M$), and manifold center correlation, which refer to dimensionality, size, and distribution of object manifolds relevant for the linear classification. In our analysis, we combine the manifold dimension and radii together into a single \textbf{manifold width} measure, defined as $R_M \cdot \sqrt{D_M}$, which captures the width of the convex hull of a manifold. Manifold width formally links the linear separability of object manifolds with their underlying geometric structure. Specifically, small values of manifold width yield more linearly separable manifold geometry. The correlation between locations of the object manifolds also plays a role in determining manifold capacity, as more correlated manifolds are less separable. For our analysis, center correlation is calculated as the average of the absolute value of the cosine similarity between pairs of object manifold centroids.


\begin{wrapfigure}{r}{0.4\textwidth}
\vspace{-4mm}
  \centering
    \begin{tikzpicture}
    \pgftext{\includegraphics[width=\columnwidth]{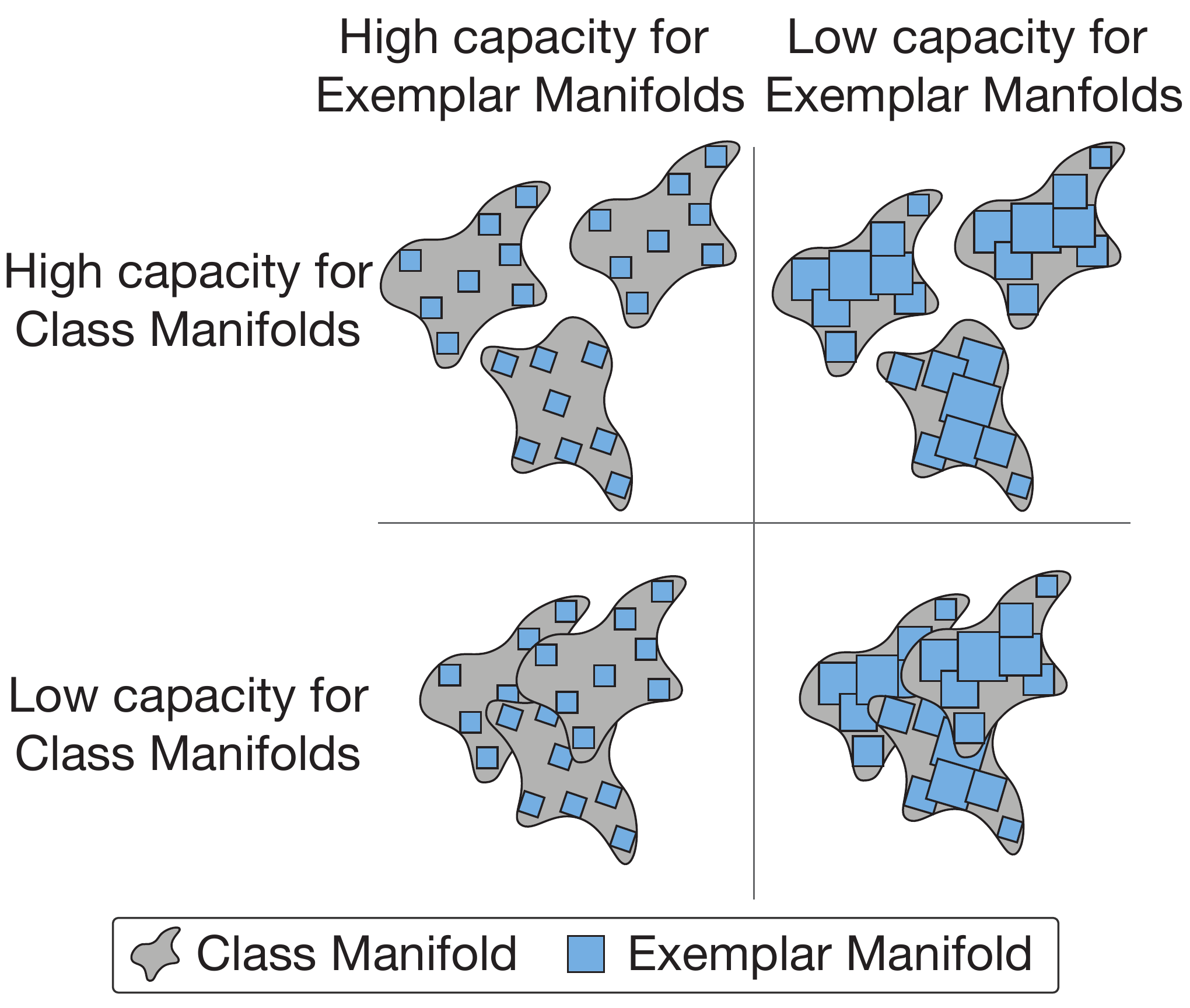}};
    \end{tikzpicture}
          \vspace{-4mm}
  \caption{\textbf{Insights from class manifolds and exemplar manifolds.} Class manifolds have variability due to differences in exemplars in the class in addition to the stochasticity or adversarial perturbations, while exemplar manifolds only have variability due to stochasticity or adversarial perturbations. \vspace{-10mm}}
  \label{fig:class_vs_exemplar}
\end{wrapfigure}

In this work, we characterize two types of manifolds:

1) \textbf{Class manifolds}: each manifold is defined by the activations evoked by multiple examples drawn from a specific class of visual or auditory stimuli (i.e., object or word identity). The variability within the manifold can come from different exemplars within the class, but may also be influenced by adversarial perturbations, and/or a layer with stochastic activations.

2) \textbf{Exemplar manifolds}: each manifold is defined by the activations evoked by multiple instances of a single exemplar (i.e., image or utterance) with the manifold variability reflecting either norm-bounded adversarial perturbations\footnote{We replicate many experiments for non-adversarial but random perturbations within the $\epsilon$-sized ball to directly compare model geometry measured on the same stimuli. Results are similar and provided in SM 6.} and/or the influence of stochasticity.

Although the class manifold analysis is most tightly coupled to the accuracy of the network on a classification task, analyzing the exemplar manifolds gives insight into how the class manifolds are constructed. Specifically, multiple types of exemplar manifold geometries can lead to the same class manifold capacity and width (Figure \ref{fig:class_vs_exemplar}).

\subsection{Quantifying object manifold overlap}
The MFTMA framework provides a geometric description of exemplar manifolds evoked by adversarial and clean stimuli in the case of multiple distributed object manifolds. However, we also characterize the overlap between adversarial and clean exemplar manifolds generated from the same stimuli (on the grounds that overlap should produce robustness to the adversarial perturbations in question). To do so, we use a notion of object manifold overlap defined as the generalization error of a linear Support Vector Machine (SVM) fit to separate two sets of representations, and report the chance-normalized error rate on the held-out data. A normalized error rate of $1$ means that two representations are completely overlapped, and $0$ means that they are completely separable and thus not overlapped. We use the \texttt{scikit-learn} SVM implementation with a train/test split of 80/20.

\subsection{Adversarial attacks}
For performing adversarial attacks, we use either the single-step fast gradient sign method (FGSM) with a random starting location or multi-step projected gradient descent (PGD). Unless otherwise specified, we use $L_\infty$ norm constrained attacks. All attacks are untargeted and performed on a model-by-model basis. When the goal is to evaluate adversarial accuracy in newly created networks with stochastic internal representations, we use ensemble-PGD, where each step is in the mean direction of k samples of the noisy gradient to ensure a useful signal \cite{Athalye2018obfuscated}. For more details, see SM 2.

\begin{wrapfigure}{r}{0.6\textwidth}
  \vspace{-6mm}
  \centering
  \includegraphics[width=\columnwidth]{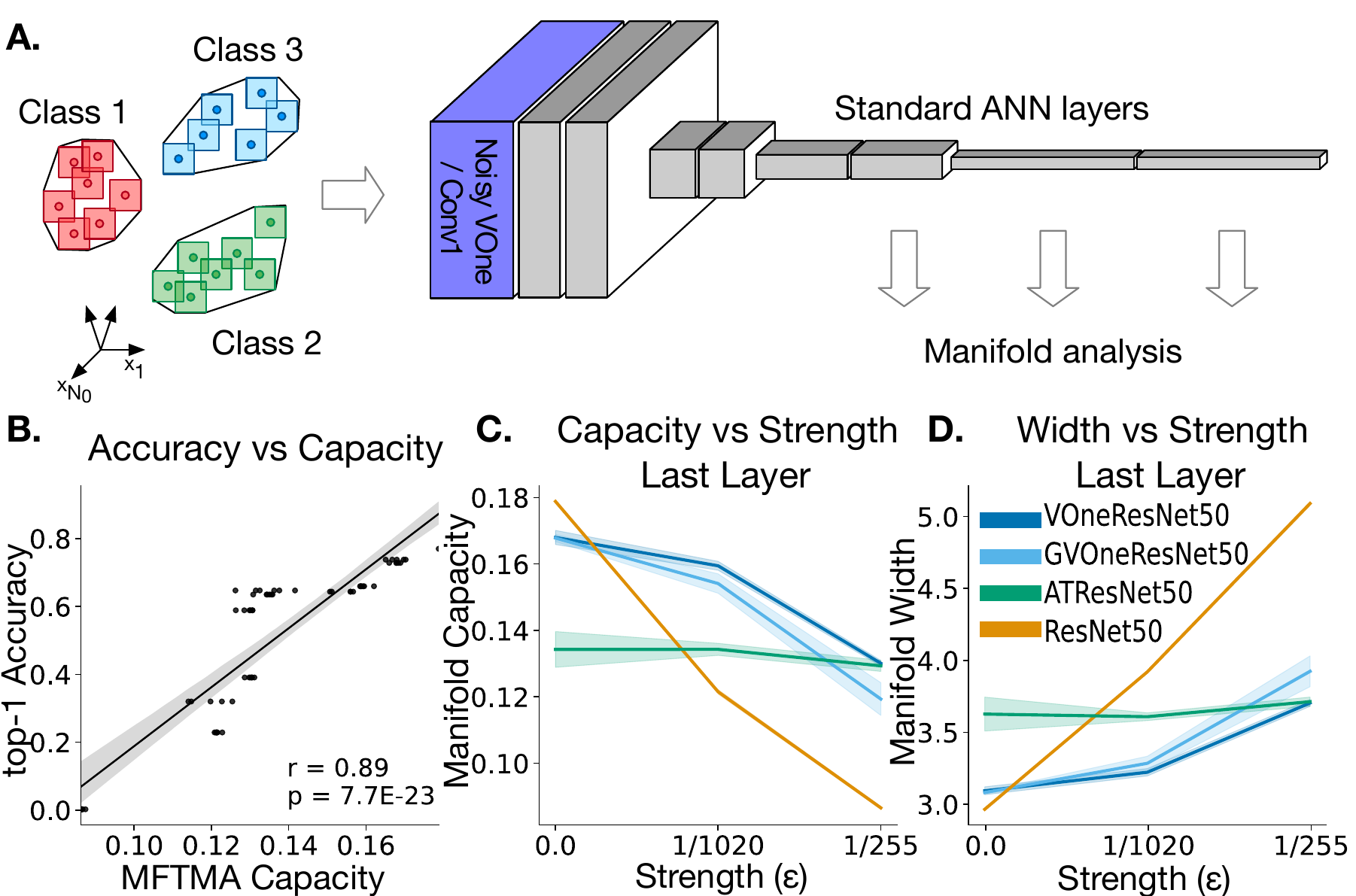}
  \vspace{-6mm}
  \caption{\textbf{Geometry and capacity of clean and adversarially perturbed class manifolds in ImageNet models} (\textbf{A}) Clean or adversarially perturbed stimuli grouped into class manifolds are provided as input to a model, internal representations are extracted, and MFTMA analysis is applied. (\textbf{B}) MFTMA capacity from the final model layer reflects accuracy from clean and adversarial images. (\textbf{C}) Manifold capacity and (\textbf{D}) manifold width for the last model layer is plotted against attack strength, revealing enhanced representational invariance in ATResNet50 (green) and to a lesser degree VOneResNet50 (dark blue) and GVOneResNet50 (light blue), while undefended ResNet50 (orange) is least invariant. Error bars are standard deviation (STD) across 5 random projection (RP) and MFTMA seeds.
  \vspace{-12mm}}
  \label{fig:vision_class_manifolds}

\end{wrapfigure}

\section{Manifold analysis of robustness in ImageNet-trained networks}\label{ImageNet}
We first investigate the properties of ImageNet-trained models, aiming to compare the geometric signatures of standard, adversarially trained, and stochastic networks.

\subsection{Models and Dataset}
We use images sampled from the ImageNet \cite{Deng2009ImageNet} test set. For class manifold analysis, the clean stimulus set consists of 50 classes, with each class containing 50 unique exemplar images for a total of 2500 unique images. For exemplar manifold analysis, 100 unique images are sampled from the ImageNet test set and each is perturbed with FGSM from a random starting location 50 times for 5000 unique images. We analyzed three publicly available ImageNet models, including ResNet50 with standard training, ResNet50 adversarially trained with an $L_\infty=4/255$ penalty (ATResNet50), and VOneResNet50, a ResNet50-based model with the first conv-relu-maxpool layers replaced by a linear-nonlinear-Poisson model front-end (called the VOneBlock) with Gabor filters and noise fitted to primate neuronal data. In addition, we analyze a novel variant of VOneResNet50 (GVOneResNet50) with additive Gaussian noise scaled to match the mean over all units in the output of the VOneBlock in response to a set of reference stimuli. GVOneResNet50 performs similarly to the original VOneResNet50 in clean and adversarial conditions (see SM 3.3 for more details). While the first two models are deterministic, VOneResNet50 and GVOneResNet50 have stochastic representations at the output of the VOneBlock. For additional details see SM 3.


\subsection{MFTMA reveals unique robustness strategies for VOneNet and adversarial training}
Our analysis begins by observing the effects of norm-bounded, gradient-based adversarial attacks on the class manifold geometry of representations in ResNet50, ATResNet50, VOneResNet50, and GVOneResNet50. For all experiments, images perturbed with PGD $L_\infty$ constraints of $\epsilon\in[0, 1/1020, 1/255]$ are shown to a model and intermediate representations are extracted, randomly projected to 5000 features \cite{Johnson1984-am, Candes2005-gb}, and analyzed with MFTMA (Figure \ref{fig:vision_class_manifolds}A). Much like previously reported with MFTMA \cite{cohen2020separability}, models develop separable representations in later layers, with capacity peaking in the final layer before classification (see SM 3.5).

First we focus on this penultimate layer, where class manifold representations have become linearly separable and where capacity thus peaks. For the full layer-wise results, see SM 3.5. We empirically confirm that capacity of the penultimate layer is predictive of the top-1 accuracy across models and attack strengths (Figure \ref{fig:vision_class_manifolds}B), indicating that MFTMA is sensitive to the attacks we are investigating. MFTMA exposes how capacity (Figure \ref{fig:vision_class_manifolds}C) and class manifold width (Figure \ref{fig:vision_class_manifolds}D) vary in the four models as a function of perturbation strength. Adversarial attacks cause the width of class manifolds to grow most rapidly in ResNet50, to a lesser degree in VOneNets, and least of all in ATResNet50. 

\begin{figure}[t!]
\vspace{-6mm}
\floatbox[{\capbeside\thisfloatsetup{capbesideposition={right,top},capbesidewidth=175pt}}]{figure}[\FBwidth]
{\caption{\textbf{Geometry and capacity of adversarially perturbed exemplar manifolds in neural networks} (\textbf{A}) Each manifold consists of a set of adversarially perturbed examples within an $\epsilon$-sized ball around a single exemplar image. (\textbf{B}) Mean capacity and manifold width as a function of layer depth; ATResNet50 maintains the size of $\epsilon$-sized ball manifolds, while undefended ResNet50, VOneResNet50, and GVOneResNet50 expand the size. (\textbf{C}) Adversarial representation capacity normalized by clean representation capacity is shown for pixels, first conv layer (the VOneBlock for VOneNets), and the last average pooling layer. VOneResNet50, GVOneResNet50, and ATResNet50 show less of a capacity decrease for low-$\epsilon$ perturbed image manifolds compared to the standard ResNet50. Error bars represent STD across 5 RP and MFTMA seeds in all plots; unnormalized capacity detailed in SM 3.6.}\label{fig:vision_eps_manifolds}}
{\includegraphics[width=210pt]{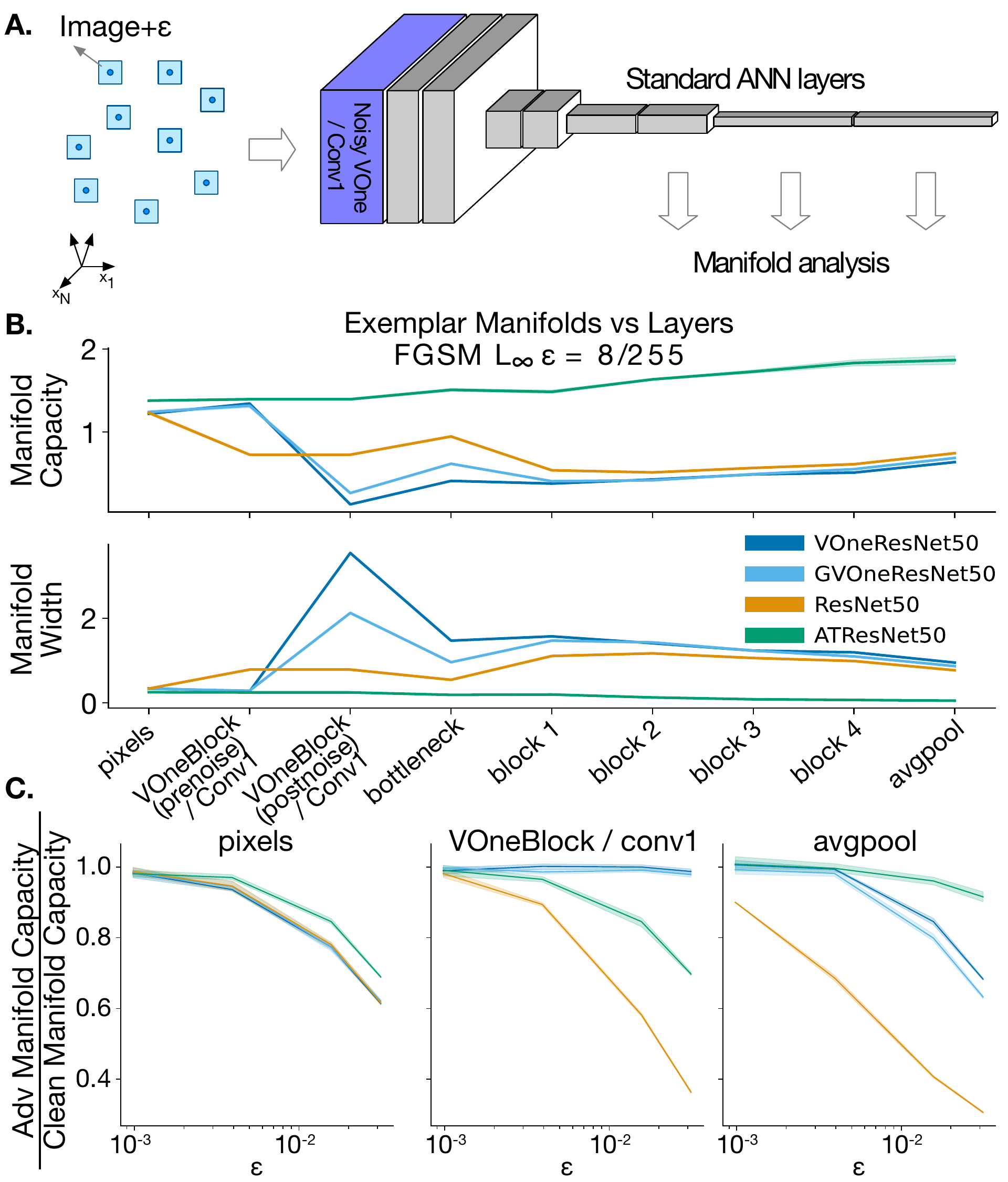}}
\end{figure}

Next, to investigate how the models represent adversarial perturbations around individual images, we introduce another MFTMA-based approach. Instead of class manifolds, we consider exemplar manifolds consisting of points sampled using FGSM with $L_\infty$ constraints of $\epsilon\in[0, 1/1020, 1/255, 4/255, 8/255]$ from a random starting location in the epsilon-sized ball around 100 exemplar images, with the goal of tracing how these $\epsilon$-sized ball exemplar manifolds develop as they travel through subsequent layers of our networks of interest (Figure \ref{fig:vision_eps_manifolds}A).

Restricting our analysis to the layer-wise trajectory of $\epsilon=8/255$ sized ball exemplar manifolds (Figure \ref{fig:vision_eps_manifolds}B), the most salient feature is how distinct the trajectories are for the ATResNet50 and VOneNets. Our results indicate the defense mechanism induced through adversarial training generally stabilizes the width of the $\epsilon$-sized ball around an exemplar as it propagates through the network, effectively mapping small perturbations around the clean image to small regions in later layers of the network. By contrast, the width of the $\epsilon$-sized ball increases in VOneNets and the standard ResNet50. In fact, at the VOneBlock output, the exemplar manifolds become highly entangled, suggesting different mechanisms of robustness for adversarially trained and stochastic models.

How then do stochastic responses improve robustness? Figure \ref{fig:vision_eps_manifolds}C demonstrates that when the capacity of the adversarially induced exemplar manifolds is normalized by the capacity for representations of the same unperturbed images (clean exemplar manifolds)\footnote{Clean exemplar manifolds for non-stochastic models have the maximum capacity of 2 \cite{chung2018classification}.}, VOneNets are far more stable than the standard ResNet50 network. At lower attack strengths where VOneNets' accuracies are minimally degraded, their normalized capacity remains stable, indicating that they represent the perturbed images with approximately the same capacity as clean images. In other words, the adversarial perturbations do not push the representation beyond that for clean images. In SM 3.7, we extend this analysis to a variety of additional networks including two networks trained adversarially with different norms, VOneResNet50 with no stochasticity during training or inference, and ResNet50 with a stochastic activation layer mirroring that in VOneResNet50 (see Dapello, Marques et al. \cite{dapello2020simulating} for network details.) Our trends hold up in each of these cases, with adversarially trained networks stabilizing the $\epsilon$-sized ball, and stochastic networks expanding it at the point of the stochastic layer, but using similar capacity relative to clean images for small perturbations. In all experiments, we find similar geometric signatures for the Gaussian and Poisson VOneNets. For simplicity in the rest of the paper we focus on stochastic representations with Gaussian noise, but include an analysis of a Poisson noise network trained on CIFAR-10 in SM 5.4.

\section{Stochastic representations improve robustness in auditory networks}\label{Auditory}

The results from the previous section highlight distinct geometric profiles for adversarially trained models and for models with stochastic responses. To demonstrate that the observed neural population geometry generalizes across modalities, we compare biologically-inspired and adversarially trained auditory models trained to perform a speech recognition task. Our focus here is not on generating an auditory model that is fully defended against adversarial attacks, but rather to test whether the influence of stochastic representations and the geometric trends observed in the previous section generalizes across domains. We thus focus on $L_\infty$ attacks acknowledging that $L_p$ audio attacks that successfully change a network's prediction are often audible to human listeners \cite{schonherr2018adversarial, qin2019imperceptible}.

\subsection{Models and dataset}
Auditory models are trained to perform the word recognition task in the Word-Speaker-Noise dataset introduced in \cite{feather2019metamers}. Networks learn to distinguish the word present in the middle of a two-second speech clip from 793 word classes. Word class manifolds are constructed from 50 unique words with 50 unique speakers saying the word, drawn from the Wall Street Journal Corpus \cite{wallstreetjournal1992} (2500 unique speech clips). Exemplar manifolds are measured from a random selection of 100 example clips from the class manifold dataset, with 50 samples measured for each clip.

\begin{figure}[b!]
  \centering
    \begin{tikzpicture}
    \pgftext{\includegraphics[width=\columnwidth]{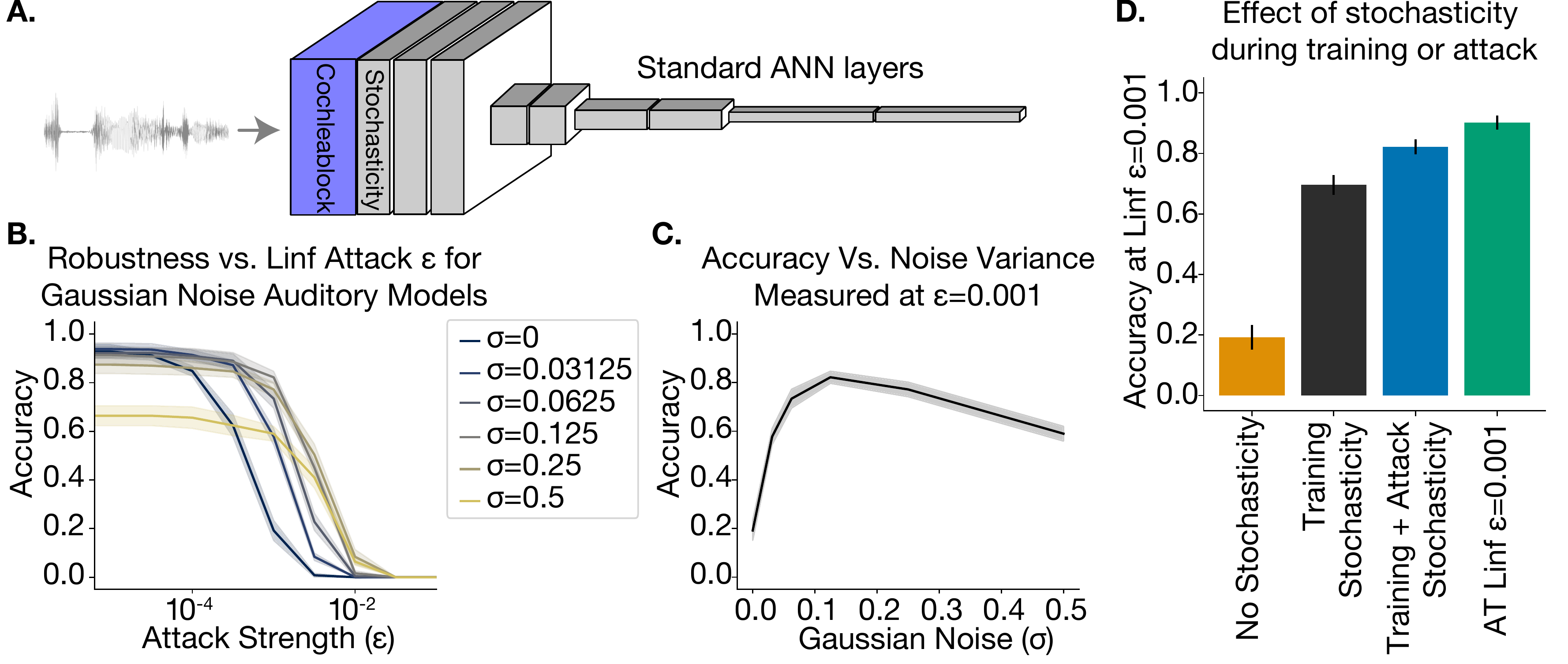}};
    \end{tikzpicture}
{\caption{\textbf{Robustness of auditory networks trained with stochastic cochlear representations} (\textbf{A}) Depiction of auditory network with stochastic cochleagram. (\textbf{B}) Performance of auditory networks trained with varying stochasticity levels evaluated on $L_\infty$ adversarial attacks, averaged over 100 randomly chosen test examples of clean speech. Error bars are STD across 5 sets of test stimuli. (\textbf{C}) An intermediate level of Gaussian noise yields the best adversarial robustness. (\textbf{D}) Stochastic activations during training but not during testing yields improvements in adversarial robustness; maintaining stochasticity during adversarial generation and testing further increases robustness.\vspace{-4mm}}\label{fig:6}}
\end{figure}

Auditory models contain a biologically-inspired `cochleagram' representation \cite{glasberg1990derivation, mcdermott2011sound}, followed by a ResNet50 architecture. The cochleagram consists of differentiable operations, allowing generation of adversarial examples in the waveform. Unlike VOneNets, and to maintain consistency with previously published auditory models \cite{kell2018task, feather2019metamers}, the architecture maintains the conv-relu-maxpoool before the first residual block, on the grounds that the cochleagram models the ear rather than primary auditory cortex. For stochastic models, Gaussian noise with standard deviation $\sigma$ is added after the cochleagram representation (Figure \ref{fig:6}A), and we refer to this model as StochCochResNet50. A comparison network without noise ($\sigma=0$) is similarly trained and evaluated (CochResNet50). A robust model is achieved with adversarial training using $L_\infty$ perturbations with $\epsilon=10^{-3}$ and maximum of 5 steps for the attack (ATCochResNet50). Training and adversarial robustness details are presented in SM 4.1 and SM 4.4 respectively.


\subsection{Adversarial robustness in auditory models with stochastic cochleagrams}
When setting the level of noise in the VOneBlock, neural responses from Macaque V1 to a specific image set were used to tune the relative amplitude of the stochastic representations. However, this type of neural data is not readily available for all sensory areas. We instead empirically found an optimal level of noise for our auditory models by varying the level of additive Gaussian noise in the stochastic cochleagram. The robustness of these models to $L_\infty$ perturbations is evaluated for different perturbation sizes. To ensure a reliable signal for adversarial attack generation, model gradients are sampled eight times for each PGD iteration. The resulting accuracy for each model is shown in Figure \ref{fig:6}B. As noise is increased, the model's robustness to adversarial perturbations increases, but only up to a point, yielding a peak in the robustness curve (Figure \ref{fig:6}C). The accuracy for StochCochResNet50 with noise during the training and attack is only slightly worse than ATCochResNet50 (Figure \ref{fig:6}D). Much like what was observed in VOneResNet50 \cite{dapello2020simulating}, including noise during training but not during adversarial evaluation significantly increases adversarial robustness, suggesting that these improvements cannot be trivially explained by the stochastic component masking the gradients during the attack. Instead, the benefits appear to reflect downstream representational changes from training with stochasticity. The additional performance boost when stochasticity is included during inference suggests a secondary defense driven by the stochastic activations during evaluation. These results show that by including stochasticity in a biologically-inspired peripheral model we can improve performance on adversarial examples and by extension learn a more human-like representation.


\begin{wrapfigure}{r}{0.67\textwidth}
\vspace{-4mm}
{\vspace{-2mm} \includegraphics[width=0.95\columnwidth]{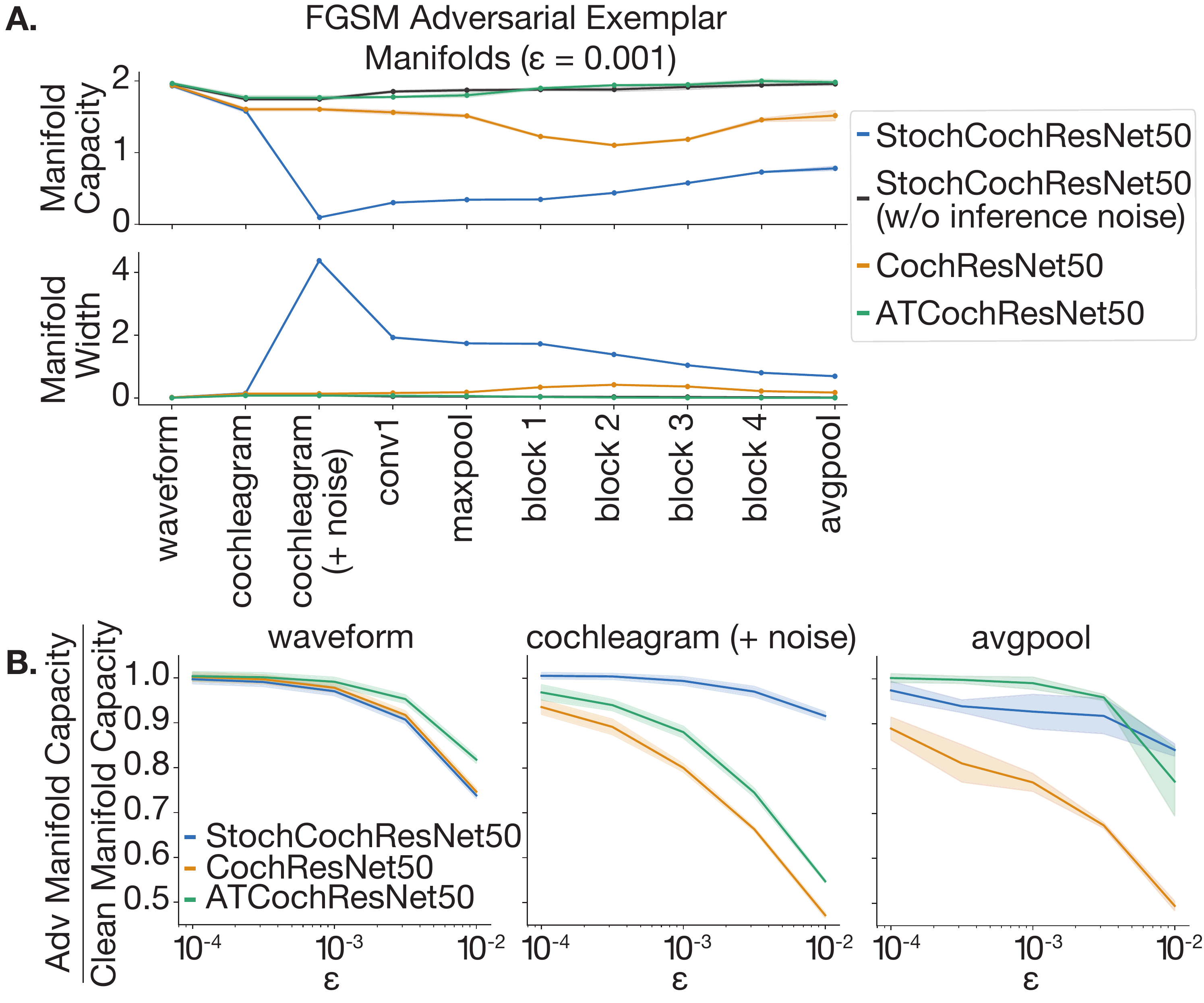}\vspace{-2mm}}
{\caption{ \textbf{Adversarial exemplar manifold geometry of auditory models.} (\textbf{A}) Capacity of adversarial exemplar manifolds for auditory models. For networks without stochasticity, the stochastic layer representation is equal to the cochleagram representation. Evaluating the capacity for the adversarial exemplar manifolds with stochasticity present during inference yields significantly lower capacity after the stochastic layer, while the model trained, but not tested, with stochasticity looks similar to the model achieved through adversarial training. Error bars are STD across 5 RP and MFTMA seeds. (\textbf{B}) Adversarial exemplar capacity normalized by the clean manifold capacity as a function of adversarial perturbation size. Unnormalized capacity detailed in SM 4.5. Error bars are STD across 5 RP and MFTMA seeds. \vspace{-6mm}}\label{fig:4_ExemplarAnalysis}}
\end{wrapfigure}

\subsection{Neural population geometry of auditory networks}
We investigated whether the neural population geometry used by StochCochResNet50 is similar to that observed in the visual domain. As with the ImageNet experiments, clean or perturbed audio is presented to the model and intermediate representations are extracted, randomly projected to 5000 features (if the layer has more than 5000 features), and analyzed with MFTMA.

An analysis of the $\epsilon$-sized ball exemplar manifolds reveals that the geometry of StochCochResNet50 with noise during training but not during evaluation is similar to that of ATCochResNet50, while the StochCochResNet50 with noise during evaluation has a much lower capacity for the adversarial exemplar manifolds at late stages of the network, much like VOneResNet50 (Figure \ref{fig:4_ExemplarAnalysis}A). As in VOneResNet50, when the $\epsilon$-sized ball exemplar manifold capacity is normalized by the clean manifold capacity, the normalized capacity for StochCochResNet50 is stable with increasing epsilon from the stochastic cochleagram layer onwards (Figure \ref{fig:4_ExemplarAnalysis}B). This suggests that StochCochResNet50 represents the adversarially perturbed sounds with the same manifold capacity and width as the clean sounds. By comparison, ATCochResNet50 does not show invariance to the adversarially perturbed stimuli until the avgpool layer.

\begin{figure}[t]
\vspace{-6mm}
{\caption{\textbf{Similarities between class manifold geometry in Auditory and ImageNet networks.} Geometry of class manifolds for Auditory (\textbf{A}) and ImageNet (\textbf{B}) networks when computed from clean or adversarial stimuli. }\label{fig:4_AuditoryAndImageGeometry}}
{\includegraphics[width=0.9\columnwidth]{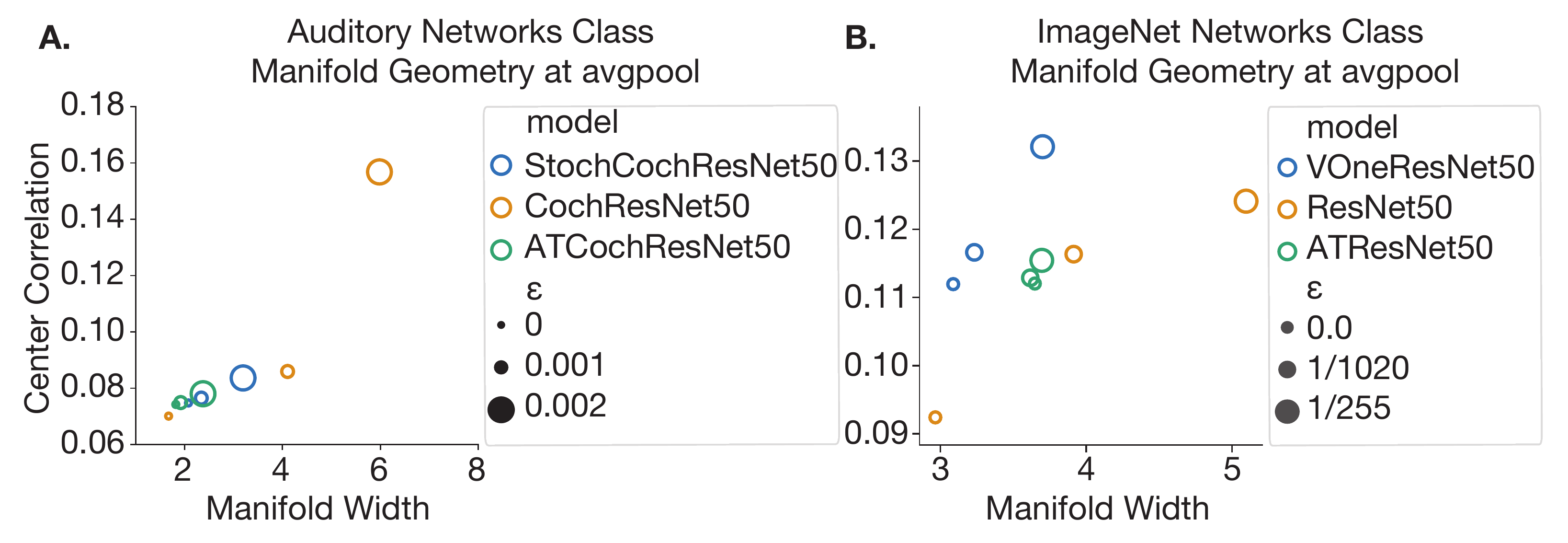}}
\end{figure}

To examine the key geometric factors associated with an increase in capacity in both the auditory and visual networks, we plot class manifold width vs. the manifold center correlation, as these two factors together determine the capacity estimate. Figure \ref{fig:4_AuditoryAndImageGeometry}A and \ref{fig:4_AuditoryAndImageGeometry}B show that in both visual and auditory networks an increase in size of the adversarial perturbation leads to an increase in manifold width and center correlation (two key variables leading to the capacity). While this general trend is present for all networks, the networks that are more adversarially robust have less change across both metrics as the perturbation size increases. This provides further evidence that the manifold metrics are a useful way to interpret the internal geometries that lead to adversarial robustness, and points to the same underlying mechanism in visual and auditory modalities for the classification degradation in the presence of adversarial vulnerability -- specifically that the class manifolds become larger and more correlated when adversarial perturbations are present.

\section{Manifold overlap and the opposing effects of noise on robust classification}\label{CIFAR}





In the previous sections we investigated how stochastic representations change the size and capacity of adversarial exemplar manifolds, but we did not directly test whether adversarial examples fall within the manifolds elicited by stochastic activations. We hypothesize that for small perturbations, activations elicited from adversarial examples overlap with the exemplar manifold formed by the stochastic representation of a clean image, effectively eliminating vulnerability to small adversarial perturbations. Here, we directly investigate whether small adversarial perturbations indeed overlap with multiple stochastic representations evoked from the same stimulus, and investigate how this overlap trades off with task performance. To do so, we test whether an SVM can separate the adversarial exemplar manifolds from the clean exemplar manifolds generated for the same stimulus. As manifold inseparability as measured by SVM is only one possible description of manifold overlap, we further detail another analysis using pairwise distance distributions in Section SM 8. Finally, we analyze the opposing effects of noise on clean and adversarial performance by investigating a CIFAR model trained with different noise levels.

\subsection{CIFAR model and datasets}
We investigate how the level of noise changes the manifold geometry in a smaller model trained on the CIFAR-10 dataset \cite{krizhevsky2009learning} with an architecture similar to ResNet18 \cite{He2015resnet}, with the first conv-relu-maxpool layers replaced by fixed-weight Gabor filters and biological-inspired activation functions adapted from the Gaussian VOneNet\footnote{More details about the model architecture and training can be found in SM 5.1}. In SM 5.4, we repeat the experiments with Poisson-like stochastic activations as well, where the trends are similar.

\subsection{Low perturbation strength adversarial examples overlap with stochastic representations}
\begin{wrapfigure}{r}{0.65\textwidth}
  \vspace{-4mm}
  \centering
    \begin{tikzpicture}
    \pgftext{\includegraphics[width=\columnwidth]{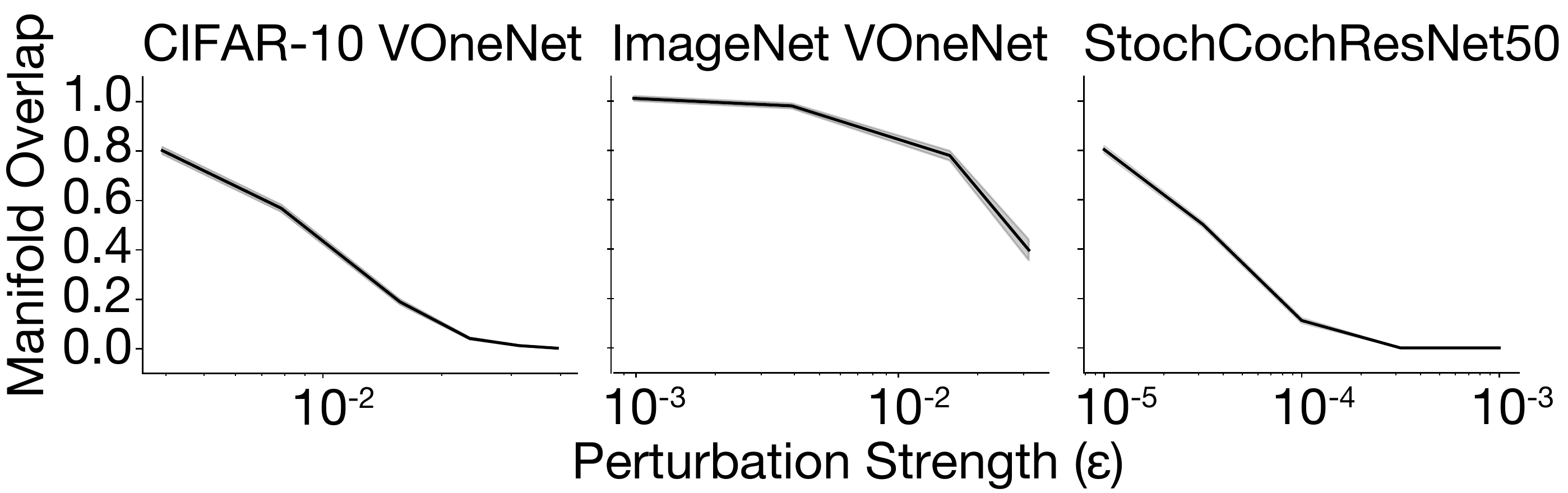}};
    \end{tikzpicture}
  \caption{\textbf{Adversarial exemplar manifolds measured at stochastic representation overlap with clean exemplar manifolds at smaller attack strengths.}
   A binary SVM trained to separate adversarial and clean exemplar manifolds extracted from the stochastic representation of each network show that clean and adversarial exemplar manifolds measured from the same stimulus become less linearly separable as the attack strength decreases. Experiments are presented for CIFAR VOneNet (Error bar is STD across 20 images and 6 RP seeds), ImageNet VOneResNet50 (Error bar is STD across 10 images and 5 RP seeds) and StochCochResNet50 (Error bar is STD across 10 images and 5 RP seeds). \vspace{-4mm}
}
  \label{fig:SVM_VS_EPS}
\end{wrapfigure}

We measure the overlap between $\epsilon$-sized ball adversarial exemplar manifolds and clean exemplar manifolds at the stochastic representations of the networks (the output of the VOneBlock for the CIFAR and ImageNet models, and the stochastic cochleagram for the auditory model). For all models, the adversarial exemplar manifold is generated by running FGSM $L_\infty$, varying the strength of the adversarial attack. The clean exemplar manifold is generated by measuring multiple stochastic representations from the same natural stimulus. We train an SVM to separate the adversarial exemplar manifold from the clean exemplar manifold (Figure \ref{fig:SVM_VS_EPS}). For CIFAR SVM experiments, the stimulus set includes 20 unique image exemplars, each with 1000 samples. For VOneResNet50 and StochCochResnet50 the stimulus set includes 10 unique stimulus exemplars, each with 5000 samples, and the features are downsampled with random projection to 5000 dimensions. In line with our hypothesis, the clean and adversarial manifolds become less overlapped as the adversarial attack strength increases and performance is degraded.

\subsection{Adversarial robustness requires balancing exemplar and class information}\label{sec:mnistoptimalnoise}
\begin{figure}[t!]
  \centering
    \begin{tikzpicture}
    \pgftext{\includegraphics[width=\columnwidth]{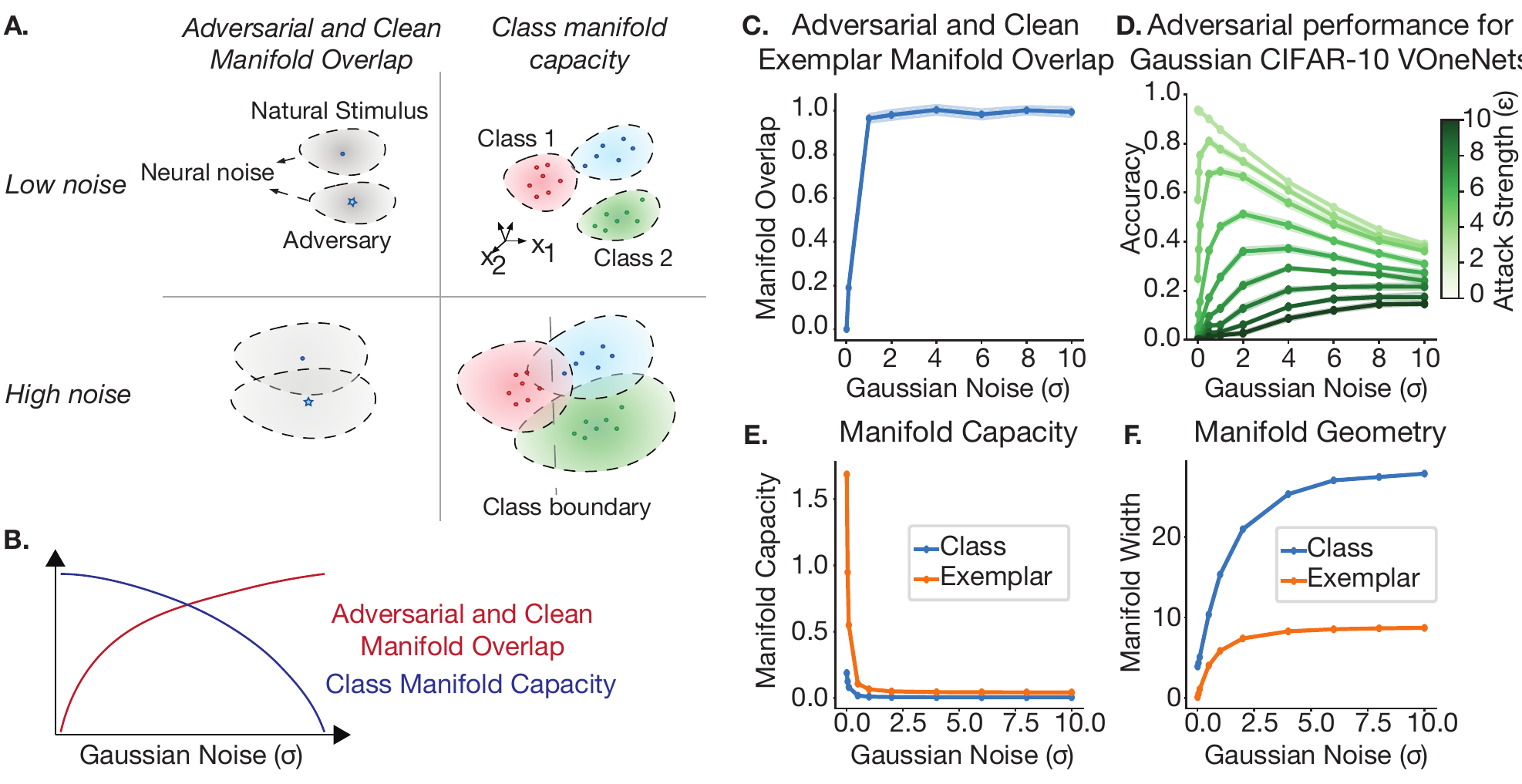}};
    \end{tikzpicture}
          \vspace{-2mm}
  \caption{\textbf{Stochastic representations induce opposing geometric effects that determine a model's performance} (\textbf{A}) Illustration of the stochasticity level's effect on the representations. In the low-noise regime, clean and adversarial representations do not overlap significantly, but class manifolds are well separated. In the high-noise regime, clean and adversarial representations are more overlapped, but the class manifolds also becomes more entangled. (\textbf{B}) These two properties trade off as the level of Gaussian noise is varied. (\textbf{C}) Binary SVMs are trained to separate adversarial and clean exemplar manifolds representations from same image in CIFAR models, showing that clean and adversarial exemplar manifolds become less linearly separable as stochasticity increases. Error bar is STD from 20 images and 6 random seeds. (\textbf{D}) Dependence between model performance and noise level at multiple attack strengths. Error bar is STD from 6 random seeds. (\textbf{E}) Capacity for both class and exemplar manifolds capacity decreases as the stochasticity increases. Error bar is STD from 6 random seeds. (\textbf{F}) Manifold width increases as the noise level increases. Error bar is STD from 6 random seeds. (All representations are from the stochastic output of the VOneBlock)}
  \label{fig:OpposingEffectsCIFAR}
  \vspace{-5mm}
\end{figure}

Although increasing the variance of stochastic representations may hide larger adversarial perturbations, high noise levels may also reduce capacity for class manifolds and decrease model accuracy. We hypothesize that the optimal noise level for model accuracy and robustness requires a balance between these two factors (Figure \ref{fig:OpposingEffectsCIFAR}A). We empirically test this hypothesis, first by showing that the adversarial and clean exemplar manifolds become less linearly separable with increasing noise variance (Figure \ref{fig:OpposingEffectsCIFAR}C). We also evaluate the performance of the CIFAR model with varying noise levels across different adversarial attack strengths (Figure \ref{fig:OpposingEffectsCIFAR}D). As the variance of the stochastic responses increase, the clean performance decreases, while performance under adversarial attack depends non-linearly on the noise variance. Using MFTMA we examine the effect of varying levels of stochasticity on class manifold capacity and related geometric properties. We generated clean exemplar and class manifolds from a stimulus set of 100 unique images, each with 50 noise samples. Figure \ref{fig:OpposingEffectsCIFAR}E shows the dependence of manifold capacity on stochasticity level. As the noise level increases, both class and exemplar manifolds become more entangled and capacity drops. In addition, Figure \ref{fig:OpposingEffectsCIFAR}F shows that as the noise level increases, the manifold width increases. Thus, both the capacity, which characterizes manifold linear separability, and the geometry demonstrate that increasing noise levels makes manifolds larger, more entangled and less linearly separable. This type of optima is also observed when choosing the level of stochasticity for the auditory networks in Figure \ref{fig:6}C, and additional auditory analysis of the noise level is found in section SM 4.6.

\section{Discussion}\label{discussion}
Using recently developed techniques to analyze the neural population geometry of a variety of networks on clean and adversarially perturbed stimuli, our work provides new insight into mechanisms underlying adversarial robustness. First, we show key geometrical differences between adversarially trained networks and VOneNets operating with stochastic neural representations. Second, we demonstrate the generality of the usefulness of stochastic representations for defending against adversarial attacks by showing that the effects extend to auditory models. It was not obvious a priori whether stochasticity would have the same benefit on auditory models in part because (unlike vision models) they are typically trained with additive noise on the input \cite{kell2018task, feather2019metamers} given that additive noise is ubiquitous in real-world auditory signals (because concurrent sounds sum together). Third, we show that stochastic representations of clean and adversarially perturbed stimuli overlap, helping to explain why the stochastic networks generalize better to perturbations below a certain threshold. Fourth, we isolate and quantify the competing effects of stochastic representations on network performance: while stochasticity increases the overlap between clean and adversarial activations, increasing stochasticity also creates more overlap between different class manifolds, making them less separable, and ultimately reducing peak clean network performance.

Here, we focused on linear methods to probe the mechanisms of adversarial robustness. Future work could benefit from more sophisticated methods to measure manifold overlap, and could also extend analysis to more naturally occurring corruptions \cite{Hendrycks2019-di}. Further, while we largely focus on the case of noise injected after fixed biologically inspired filters, more work is needed to explore what types of representations benefit from the addition of noise; for instance Dapello, Marques et al. found there was not as strong of a protective effect with stochasticity added to a standard convolutional filter. We leave the exact role of the neurally plausible filters, as used in both VOneResNet50 and StochCochRenet50, as a promising direction for future work. The presence of noise at all stages of the brain, suggests that if we can resolve when stochasticity is useful, this defense may be extensible to additional network layers for greater gains in robustness.

In theoretical neuroscience, much of the discussion on the role of stochasticity in neural coding has centered around the efficient representation of uncertainties associated with task-relevant variables \cite{ma2006bayesian, ma2014neural}. Our work adds to this line of research, by using geometry to demonstrate how stochasticity improves the neural population's robustness to adversarial perturbations unseen during training in deep networks. We hope that our work will motivate mechanistic explanations of biologically plausible robust computation in ANNs through the lens of geometry, and further identification of biological constraints that might inspire favorable changes in task-efficient neural representations.

\everypar{\looseness=-1}

\begin{ack}
This work was supported in part by NSF Neuronex 1707398 (S.C.), the Gatsby Charitable Foundation GAT3708 (S.C.), BCS Computational Fellowship (S.C.), Intel Research Grant (S.C., H.L.), a Friends of the McGovern Institute Fellowship (J.J.F.), a DOE CSGF Fellowship (J.J.F.), the PhRMA Foundation Postdoctoral Fellowship in Informatics (T.M), the Semiconductor Research Corporation (SRC) (J.D., J.J.D.), NIH grant R01DC017970 (J.H.M.), the Office of Naval Research grant MURI-114407 (J.J.D.), the Simons Foundation grant SCGB-542965 (J.J.D.), the MIT-IBM Watson AI Lab grant W1771646 (J.D, J.J.D.). All experiments were performed on the MIT BCS OpenMind Computing Cluster.
\end{ack}
\bibliographystyle{unsrt}
\bibliography{adv_manifolds}

\end{document}


\textbf{\Large Supplementary Material} 

\section{Manifolds analysis}
\subsection{Replica-based Mean Field Theory Manifold Analysis}
In this section, we provide a more complete description of Replica-based Mean-Field Theoretic Manifold Analysis (MFTMA). MFTMA refers to the method that was first introduced in \cite{chung2018classification}, and this framework has been used to analyze internal representations of deep networks, ranging from visual \cite{cohen2020separability} to speech \cite{stephenson2019untangling} and natural language tasks \cite{mamou2020emergence}. The shorthand MFTMA was first used in \cite{stephenson2019untangling}, and provided code formed the basis for our analysis\footnote{\href{https://github.com/schung039/neural_manifolds_replicaMFT}{https://github.com/schung039/neural\_manifolds\_replicaMFT}}.

As noted in the main text, object manifolds as used within MFTMA are defined as a set of stimulus evoked representations, grouped by categorical labels. Examples of object manifolds used in this work are population responses to different exemplars in the same object class (a class manifold), or an exemplar manifold created by a variability around a single stimulus (i.e, a single image or an utterance), where variability in the manifold can be from an adversarial perturbation or stochasticity.

Derived using replica theory in statistical physics, the MFTMA framework generalizes the theory of perceptron classification capacity for discrete point patterns \cite{gardner1988space} to the capacity of object manifolds. Specifically, the MFTMA framework measures the manifold capacity, defined as the maximum number of object manifolds such that the majority of the ensemble of random dichotomy labels for these objects manifolds can be linearly separated. This is a direct generalization of `shattering' capacity of a perceptron, where the counting unit for the perceptron is the number of objects \footnote{where each object's manifold can include a finite or infinite number of points}, rather than number of discrete patterns. As the measure of manifold capacity can be empirically evaluated (just as the perceptron capacity can be empirically evaluated), the match between the empirical manifold capacity and the theoretical manifold capacity predicted from the object manifold properties has been shown in many domains with different datasets \cite{chung2018classification, cohen2020separability, stephenson2019untangling, mamou2020emergence}. The expression for manifold capacity in the MFTMA framework gives rise to new measures for characterizing geometric properties of object manifolds, such that the shattering capacity of object manifolds can be formally expressed in terms of the geometric properties of object manifolds.  As the framework formally connects the representational geometric properties and the object manifold's classification capacity, the measures from this framework are particularly useful for gaining a mechanistic account of how information content about objects are embedded in the structure of the internal representations from deep networks. Below we provide additional details of the measures from this framework: manifold capacity and the geometrical properties (such as manifold dimension, radius, width, and manifold center correlation).

\subsection{Metrics in the Manifold Capacity Theory}
Given neural or feature representations where $P$ object manifolds are embedded in $N$-dimensional ambient feature (or neural state) space, \textit{load} is defined as $P/N$. Large/small load implies that many/few object manifolds are linearly separable in the feature dimension. Consider a linear classification problem where binary positive and negative labels are assigned randomly to $P$ object manifolds, while all the points within the same manifold share the same label, and the problem is to find a linearly classifying hyperplane for these random manifold dichotomies.

\paragraph{Manifold capacity} is defined as the critical load $\alpha_M = P / N$ such that above this value, most dichotomies have a linearly separating solution, and below this value, most of the dichotomies do not have a linearly separating solution. A system with a large manifold capacity has object manifolds that are well separated in the feature space, and a system with a small manifold capacity has object manifolds that are highly entangled (ie, not linearly separable) in the feature space.


Manifold capacity can be estimated using the replica mean field formalism with the framework introduced by \cite{chung2018classification} and refined in \cite{cohen2020separability}. As mentioned in the main text, $\alpha_{M}$ is estimated as $\alpha_{MFT}$, or MFTMA manifold capacity, from the statistics of \textit{anchor points}, $\tilde{s}$, a representative point for the points within an object manifold that contributes to a linear classification solution. The general form of the MFTMA manifold capacity has been shown \cite{chung2018classification, cohen2020separability} to be:

$$\alpha_{MFT}^{-1}=\left\langle \frac{\left[t_{0}+\vec{t}\cdot\tilde{s}(\vec{t})\right]_{+}^{2}}{1+\left\Vert \tilde{s}(\vec{t})\right\Vert ^{2}}\right\rangle _{\vec{t},t_0}$$

where $\left\langle \ldots\right\rangle _{\vec{t},t_0}$ is a mean over
random $D$- and 1- dimensional Gaussian vectors $\vec{t}, t_0$ whose components are i.i.d. normally distributed $t_{i}\sim\mathcal{N}(0,1)$.

This framework introduces the notion of \textit{anchor points}, $\tilde{s}$, uniquely given by each $\vec{t}$, a coordinate for the manifold's embedded space, and $t_0$, the manifold's center direction, representing the variability introduced by all other object manifolds, in their arbitrary orientations. Formally, $\tilde{s}$ represents a weighted sum of support vectors contributing to the linearly separating hyperplane in KKT (Karush–Kuhn–Tucker) interpretation \cite{chung2018classification}.

Formalized in this way, manifold capacity has many useful interpretations. First, as the manifold capacity is defined as the critical load for a linear classification task, it captures the linear separability of object manifolds. Second, the manifold capacity is defined as the \textit{maximum} number of object manifolds that can be packed in the feature space such that they are linearly separable, it has a meaning of how many object manifolds can be "stored" in a given representation such that they can distinguished by the downstream linear readout. Third, the manifold capacity captures the amount of linearly decodable object information per feature (or neuron) dimension embedded in the distributed representation.

\subsection{Manifold Geometric Measures}
The statistics of the anchor points play a key role in estimating a object manifold's effective Manifold Radius $R_M$ and Manifold Dimension $D_M$, as they are defined as:

$$R_{\text{M}} =\sqrt{\left\langle \left\Vert \tilde{s}(\vec{T})\right\Vert ^{2}\right\rangle _{\vec{T}}}$$

$$D_{\text{M}}=\left\langle \left(\vec{t}\cdot\hat{s}(\vec{T})\right)^{2}\right\rangle _{\vec{T}}$$

where $\hat{s} = \tilde{s} / \Vert \tilde{s} \Vert$ is a unit vector in the direction of $\tilde{s}$, and $\vec{T}=(\vec{t},t_0)$.

\paragraph{Manifold Dimension} measures the effective dimension of the projection of $\vec{t}$ on its unique anchor point $\tilde{s}$, capturing the dimensionality of the regions of the manifolds playing the role of support vectors. In other words, the manifold dimension is the dimensionality of the object manifolds realized by the linearly separating hyperplane. High values of $D_M$ imply that the fraction of the part within the object manifold embedded in the margin hyperplane is high-dimensional, thereby implying that the classification problem is hard.

\paragraph{Manifold Radius} measures the average norm of the anchor points, $\tilde{s}(\vec{T})$, capturing the size of the object manifold realized by the linearly separating hyperplane. A small value of $R_M$ implies tightly grouped anchor points.

\paragraph{Manifold Width} combines the two measures contributing to the manifold's overall width in the dimensional space, namely, Manifold Radius, $R_M$, and Manifold Dimension, $D_M$. Prior theoretical work has shown that there is a trade-off between $R_M$ and $D_M$ such that as long as $R_M\cdot \sqrt{D_M}$ stays constant, the manifold capacity stays constant \cite{chung2016linear, chung2018classification}.

If the object manifold centers are in random locations and orientations, these geometric properties predict the MFTMA manifold capacity \cite{chung2018classification}, by

$$\alpha_{\text{MFT}}\approx\alpha_{\text{Ball}}\left(R_{\text{M}},\,D_{\text{M}}\right)\approx\alpha_{\text{point}}\left(R_{\text{M}}\cdot \sqrt{D_{\text{M}}}\right)$$
where,

$$\alpha_{\text{Ball}}^{-1}(R,D)=\int_{-\infty}^{R\sqrt{D}}Dt_{0}\frac{(R\sqrt{D}-t_{0})^{2}}{R^{2}+1}$$
is a capacity of $L_2$ spheres with radius $R$ and dimension $D$ as defined in \cite{chung2016linear} and
$$\alpha_{\text{Point}}^{-1}(\kappa)=\int_{-\infty}^{\kappa}Dt(t-\kappa)^2 $$

is a classification capacity of points given an imposed margin of $\kappa$ as defined in \cite{gardner1988space,chung2018classification}.

In real data, the manifolds have various correlations, hence the above formalism has been applied to the data projected into the null spaces of manifold centers, similar to the method proposed by \cite{cohen2020separability}. To characterize the correlation structure in the data, we compute average of absolute values of pairwise cosine correlation between given manifolds' centroids, and provide them alongside other geometric measures here in SM.

\subsection{Capacity: theory vs. simulation}
While a good match between empirically observed manifold capacity ($\alpha_M$)\footnote{The empirical capacity is computed using a bisection search on the critical number of feature dimensions required in order to reach roughly 50\% chance of having linearly separable solutions given a fixed number of manifolds and their geometries} and MFTMA predicted manifold capacity ($\alpha_{MFT}$) has been demonstrated in numerous past works \cite{cohen2020separability, stephenson2019untangling, mamou2020emergence}, here we verify the match between predicted and simulated capacity under adversarial conditions.

\begin{figure}[h!]
  \centering
    \begin{tikzpicture}
    \pgftext{\includegraphics[width=0.6\columnwidth]{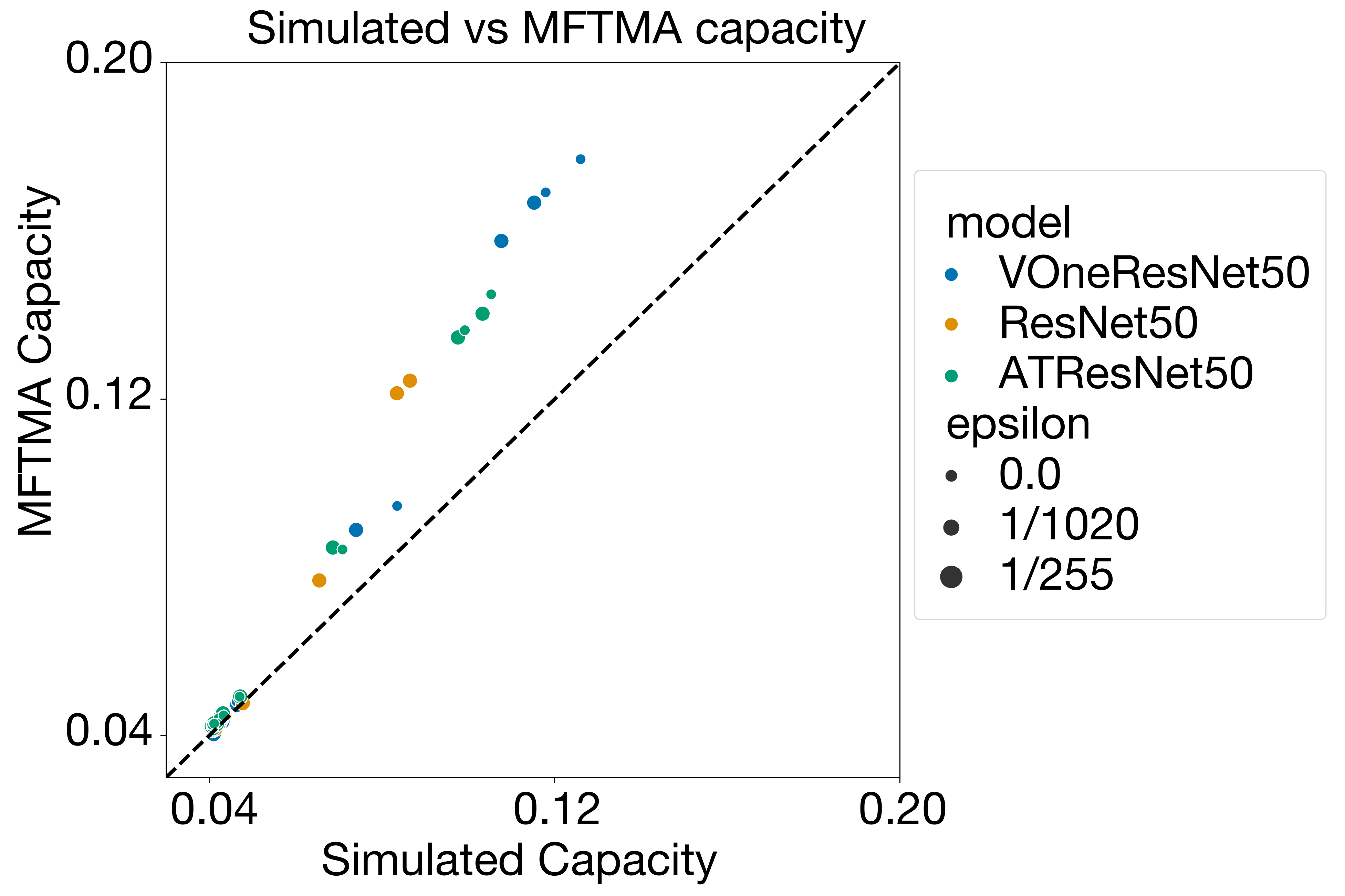}};
    \end{tikzpicture}
          \vspace{-4mm}
  \caption{\textbf{MFTMA manifold capacity approximately predicts empirically observed (simulated) manifold capacity under clean and adversarial conditions} for all layers investigated in VOneResNet50 (blue), ResNet50 (orange), and ATResNet50 (green) and across different strengths of adversarial attack, consistent with prior work \cite{cohen2020separability}.}
  \label{fig:simcap_vs_mftma}
\end{figure}

In Figure \ref{fig:simcap_vs_mftma} we show empirical (simulated) capacity vs MFTMA predicted capacity for class manifolds in all analyzed layers of VOneResNet50, ResNet50, and ATResNet50 and under a variety of adversarial perturbation strengths. We observe a tight relationship between MFTMA and simulated capacity, with a mild propensity for MFTMA to overestimate capacity as representations become more separable, consistent with prior observations \cite{cohen2020separability}, indicating that MFTMA is also applicable when representations arise from adversarial stimuli.

\section{Adversarial Attacks}\label{sec:adv_attacks}
For performing white box adversarial attacks, we used the single-step fast gradient sign method (FGSM) \cite{Szegedy2013intriguing} or multi-step projected gradient descent (PGD) \cite{Madry2017advtrain} with an $L_{p}$ norm constraint. Given an image or waveform $x$, These method uses the gradient of the loss to construct an adversarial image $x_{adv}$ which maximizes the model loss within an $L_p$ bound around $x$. Unless stated otherwise, we use $L_\infty$ attack constraints. Formally, an $L_\infty$ PGD attack iteratively computes $x_{adv}$ as

$$x^{t+1} = \mathsf{Proj}_{x+\mathcal{S}}(x^{t}+\alpha\, \mathsf{sgn}(\nabla_{x^{t}}L(\theta,x^{t},y)))$$

where $x$ is the original image or audio signal, and the $Proj$ operator ensures the final computed adversarial image $x_{adv}$ is constrained to the space $x+\mathcal{S}$, in this work the $L_{\infty}$ ball around $x$. FGSM is a special case of PGD, where only a single step is taken.

We used FGSM with a random starting location in the $\epsilon$-sized ball exemplar manifold experiments because we found that when using 64 step PGD for sampling 50 locations around each image, we frequently recovered the same perturbation multiple times, in particular for adversarially trained networks. Therefore, we resorted to FGSM with random starting locations in order to ensure a diversity of sample points in the $\epsilon$-sized ball around our exemplars.

Model specific details of adversarial attacks are provided in each model section below.

\section{ImageNet vision networks}\label{sec:imagenet_vis_networks}

\subsection{Model architecture and training details}
With the exception of the Gaussian VOneResNet50, (GVOneResNet50), all of the ImageNet \cite{Deng2009ImageNet} trained models investigated including VOneResNet50, ResNet50, and ATResNet50 were drawn from publicly available sources. ResNet50 pre-trained on ImageNet was taken from \url{https://pytorch.org/vision/0.8/models.html}. ATResNet50 (adversarially trained on ImageNet with PGD $L_\infty$ $\epsilon=4/255$) was taken from \url{https://github.com/MadryLab/robustness} \cite{robustnesslib}. VOneResNet50 was taken from \url{https://github.com/dicarlolab/vonenet}. ResNet50 and ATResNet50 share the same architecture, as described in \cite{He2015resnet}. VOneResNet50 and GVOneResNet50 have the first conv-relu-maxpool of a ResNet50 architecture replaced with a linear-nonlinear model, followed by Poisson-like or Gaussian noise respectively (VOneBlock). More specifically, the VOneBlock consists of a Gabor filter bank tuned to match primate primary visual cortex neuronal data, simple and complex cell non-linearities, and a stochastic layer. The stochastic layer of GVOneResNet50 was composed of zero-mean Gaussian noise with the standard deviation matched to the overall mean across all VOneBlock units in response to a reference stimulus set of natural images. The stochastic layer of VOneResNet was a continuous, second order approximation of Poisson noise as described in Dapello, Marques et al. All components of VOneResNet50 and GVOneResNet50 including the VOneBlock are fully differentiable, and the models are always adversarially attacked end-to-end. For more details on VOneNets, we refer the reader the main text and supplemental materials of \cite{dapello2020simulating}.

GVOneResNet50 was created using the public \url{https://github.com/dicarlolab/vonenet} repository. Like existing VOneNets, GVOneResNet50 was trained on ImageNet, with standard preprocessing and data augmentation during training including random resizing and cropping to $224 \times 224$ pixels and random horizontal flipping. For validation, images are center cropped to $224 \times 224$ pixels. Preprocessing was followed by a normalization to render all pixel values between 1 and -1. We used a batch size of 256 images and trained on 2 QuadroRTX6000 GPUs for 70 epochs on the MIT BCS OpenMind computing cluster, for a total training time of approximately 80 hours. We used a step learning rate schedule with 0.1 starting learning rate, divided by 10 every 20 epochs with Stochastic Gradient Descent, a weight decay 0.0001, momentum 0.9, and cross-entropy loss between image labels and model predictions (logits).

Intermediate layers selected for analysis from ResNet50 and ATResNet50 include the pixels, the first conv-relu (Conv1), the maxpool (bottleneck), the output of each residual block (block1, block2, block3, block4), and the average pooling layer (avgpool) prior to the softmax classification layer. For VOneResNet50, and GVOneResNet50 selected layers include pixels, the VOneBlock before the stochastic layer and after the stochastic layer, the $1\times1$ convolution following the VOneBlock (bottleneck), the output of each residual block (block1, block2, block3, block4), and the average pooling layer (avgpool) layer prior to the softmax classification layer. To make layerwise plots well aligned, for ResNet50 and ATResNet50, the Conv1 layer is duplicated at the pre and post noise points in the trajectory, essentially plotting them as if they had a noise layer with 0 variance following Conv1.

\subsection{Adversarial attacks}

For ImageNet models (VOneResNet50, GVOneResNet50, ResNet50, and ATResNet50), attacks on class manifold images were performed with untargeted PGD, using 64 steps and a step size of $\epsilon/8$. When evaluating network accuracy to adversarial attacks we average over 8 gradient samples at each step of optimization to ensure useful information from stochastic gradients \cite{Athalye2018obfuscated}. We did not average gradients across multiple noise samples for the MFTMA experiments on stochastic ImageNet models, because we are not strictly focused on claims of robust accuracy under a worst case attack. For adversarial $\epsilon$-sized ball exemplar manifolds, FGSM with a random starting location was used, and we verified that all 50 sample points generated were unique images. All adversarial attacks for ImageNet models were performed using the adversarial robustness toolbox \cite{art2018}.

\subsection{Characterizing adversarial robustness of ImageNet models}
\begin{figure}[h!]
  \centering
    \begin{tikzpicture}
    \pgftext{\includegraphics[origin=c,width=0.9\columnwidth]{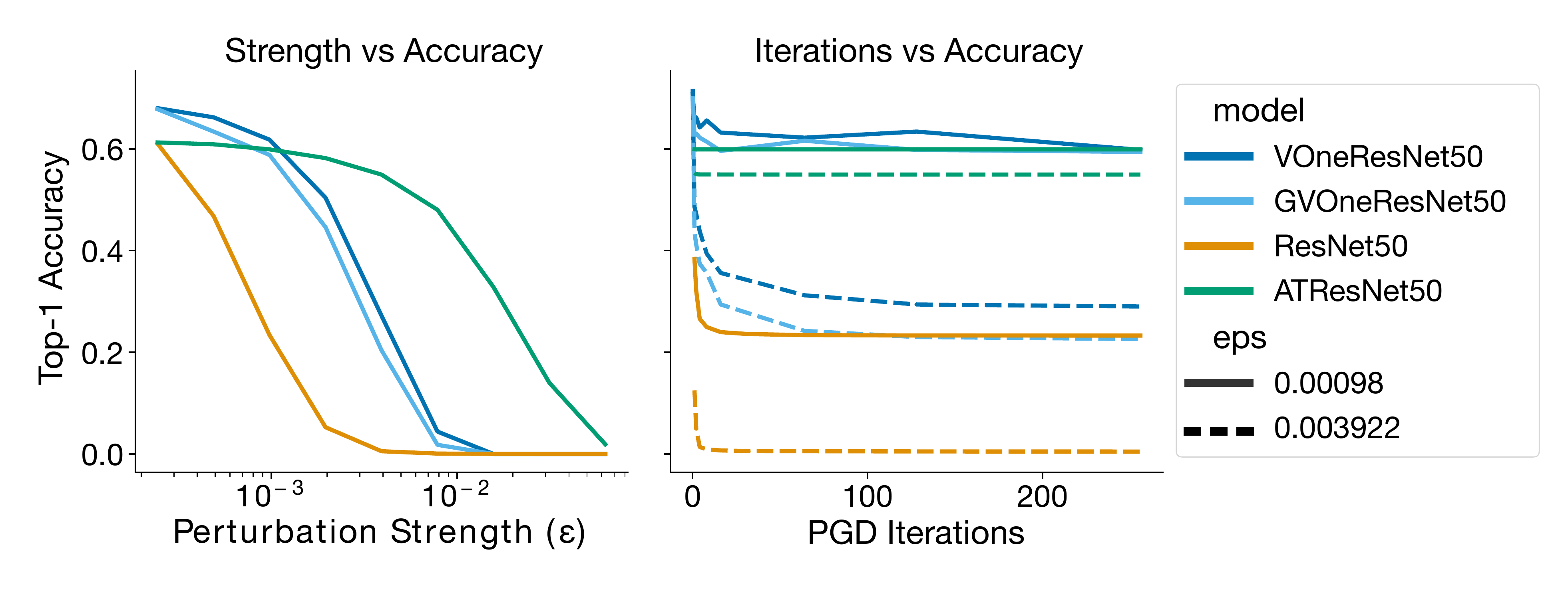}};
    \end{tikzpicture}
          \vspace{-4mm}
  \caption{
  \textbf{Adversarial robustness of VOneResNet50, GVOneResNet50, ATResNet50 and ResNet50.} Left: the $L_\infty$ PGD perturbation strength ($\epsilon$) vs top-1 accuracy on 5000 ImageNet validation images for VOneResNet50, GVOneResNet50, ATResNet50 and ResNet50. While the original Poisson-like noise VOneResNet50 is slightly more robust than GVOneResNet50, both models are in a comparable range. Right: PGD iterations vs ImageNet top-1 accuracy curve going from zero (random perturbation) to one to many shows that the gradients used in our PGD attack for both models contain useful information for computing adversarial attacks.}
  \label{fig:characterizing_GVOne}
\end{figure}

Here, we characterize the adversarial robustness of the VOneResNet50, GVOneResNet50, ATResNet50, and ResNet50. Of particular interest is GVOneResNet50 in comparison to the original VOneResNet50. Figure \ref{fig:characterizing_GVOne} shows the strength accuracy curves of VOneResNet50 compared to GVOneResNet50 for an untargetted $L_\infty$-constrained Projected Gradient Descent (PGD) attack with 64 PGD iterations, a step size of $\epsilon/8$, and eight gradient samples at every step. The plot demonstrates that the GVOneResNet50 is only marginally less robust than the original Poisson-like noise VOneResNet50, indicating that Poisson-like stochasticity is not necessary for improvements in adversarial robustness.

Figure \ref{fig:characterizing_GVOne} also provides a number of useful sanity checks on the validity of our attacks for these models. For all models, accuracy clearly transitions smoothly from near clean level performance to $0\%$ accuracy as strength increases, indicating that gradients include useful signal for computing optimal image perturbations. Furthermore, increasing the number of gradient iterations from zero (a random starting point) to one and again to many iterations generally increases the effectiveness of the attack, again demonstrating the quality of the gradients for computing adversarial attacks. In general, we did not find multiple random starting points or increasing the number of gradient samples per step beyond eight to have any significant effect on model accuracy.

\subsection{Accuracy vs. capacity}
\begin{figure}[h!]
  \centering
    \begin{tikzpicture}
    \pgftext{\includegraphics[width=0.6\columnwidth]{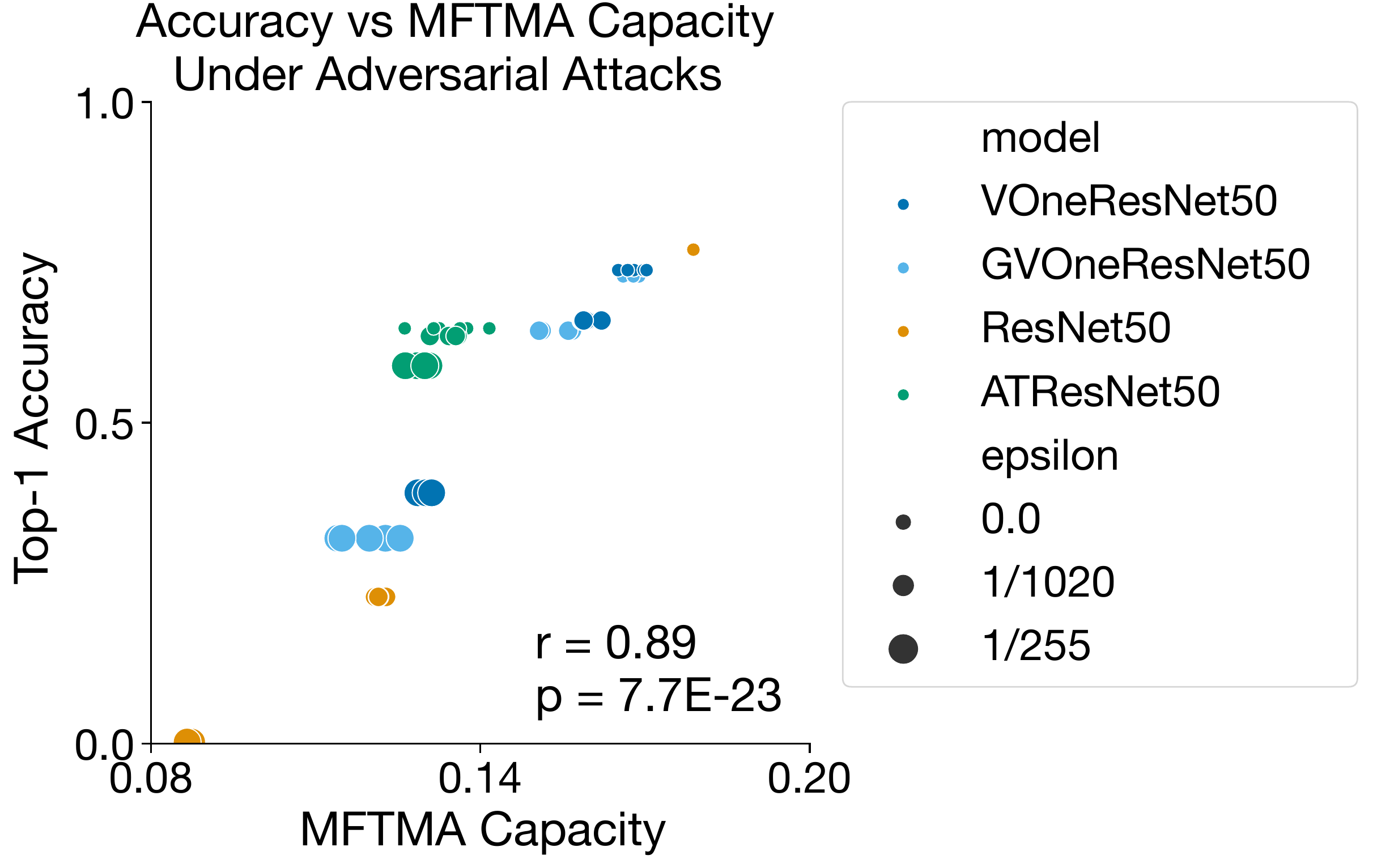}};
    \end{tikzpicture}
          \vspace{-4mm}
  \caption{\textbf{MFTMA predicted capacity of class manifolds is well correlated with a model's performance under adversarial attack.} Models include VOneResNet50 (blue), GVOneResNet50 (light blue), undefended ResNet50 (orange), and ATResNet50 (green) for clean images and PGD $L_\infty$ $\epsilon=1/1020$ and $\epsilon=1/255$ perturbed images.}
  \label{fig:cap_vs_acc}
\end{figure}

In Figure \ref{fig:cap_vs_acc} we include a more detailed form of Figure 2B, depicting how models and attack strength conditions tested for adversarial accuracy relate to MFTMA capacity.

\subsection{Full layerwise trajectories of class manifold geometry}
\begin{figure}[h!]
  \centering
    \begin{tikzpicture}
    \pgftext{\includegraphics[width=0.7\columnwidth]{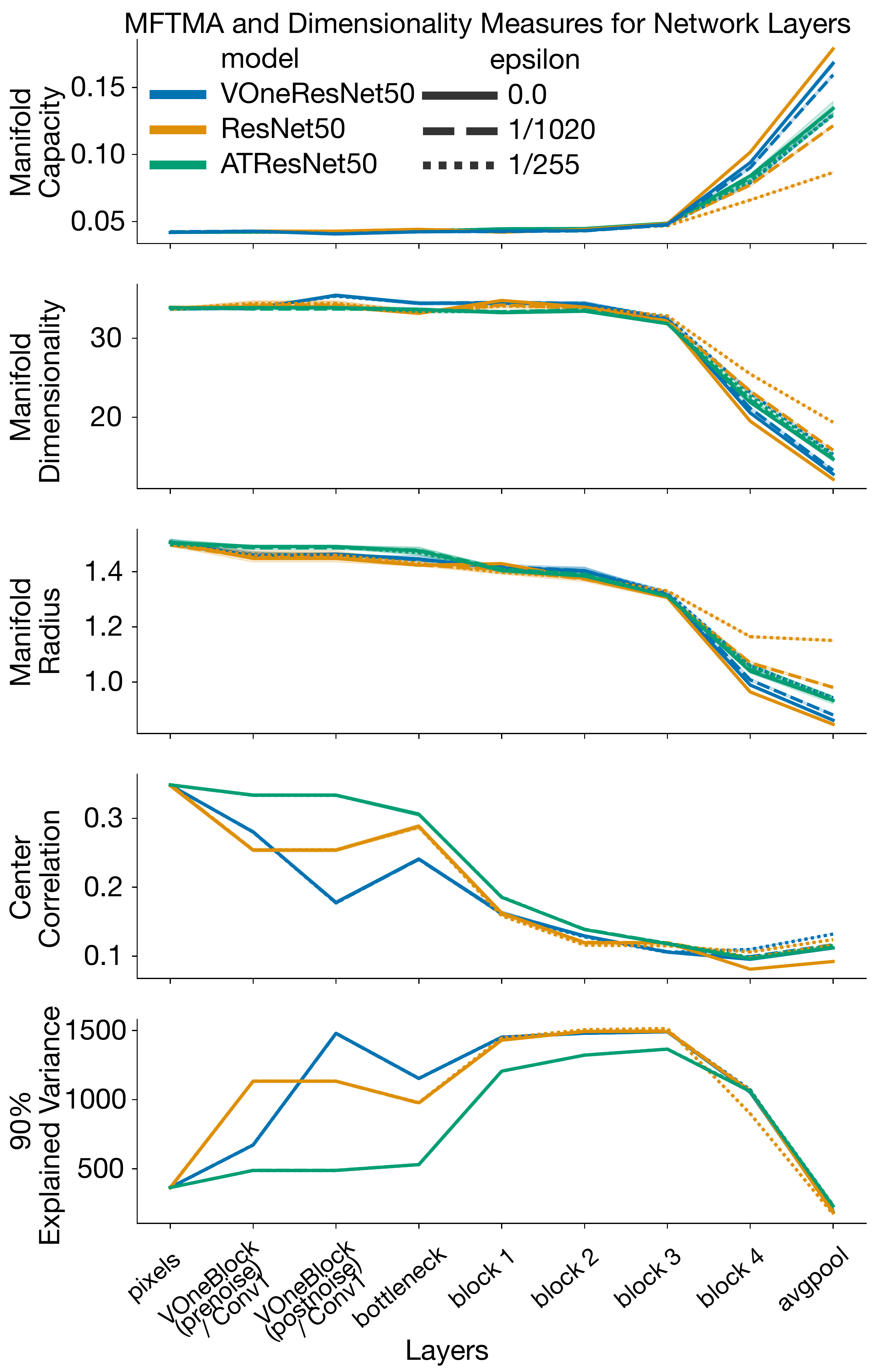}};
    \end{tikzpicture}
          \vspace{-4mm}
  \caption{\textbf{Class manifold measures across layers} for VOneResNet50 (blue), ResNet50 (orange), and ATResNet50 (green) across different attack strengths. From top to bottom: mean manifold capacity, manifold dimension, manifold radius, manifold center-center correlation, and the number of principal components need to retain 90\% of the total data variance. Error bars represent standard deviation (STD) for 5 RP and MFTMA seeds.}
  \label{fig:VOneNet_all_layers_all_measures}
\end{figure}

Figure \ref{fig:VOneNet_all_layers_all_measures} extends the class manifold analysis performed in Figure 2 to show the MFTMA and dimensionality measures from all network stages of VOneResNet50, ResNet50, and ATResNet50. Manifold capacity, $R_M$, $D_M$, center correlation, and the number of principal components needed to capture $90\%$ of the variance were all measured from the class manifold dataset in used in Figure 2. As expected, the class manifold capacity increases in deeper layers of the network and is highest at the final stage of the network (avgpool) for all tested networks.

\subsection{Exemplar manifold capacity for different adversarial strengths}\label{sec:unnormvisioncap}
\begin{figure}[b!]
  \centering
    \begin{tikzpicture}
    \pgftext{\includegraphics[width=\columnwidth]{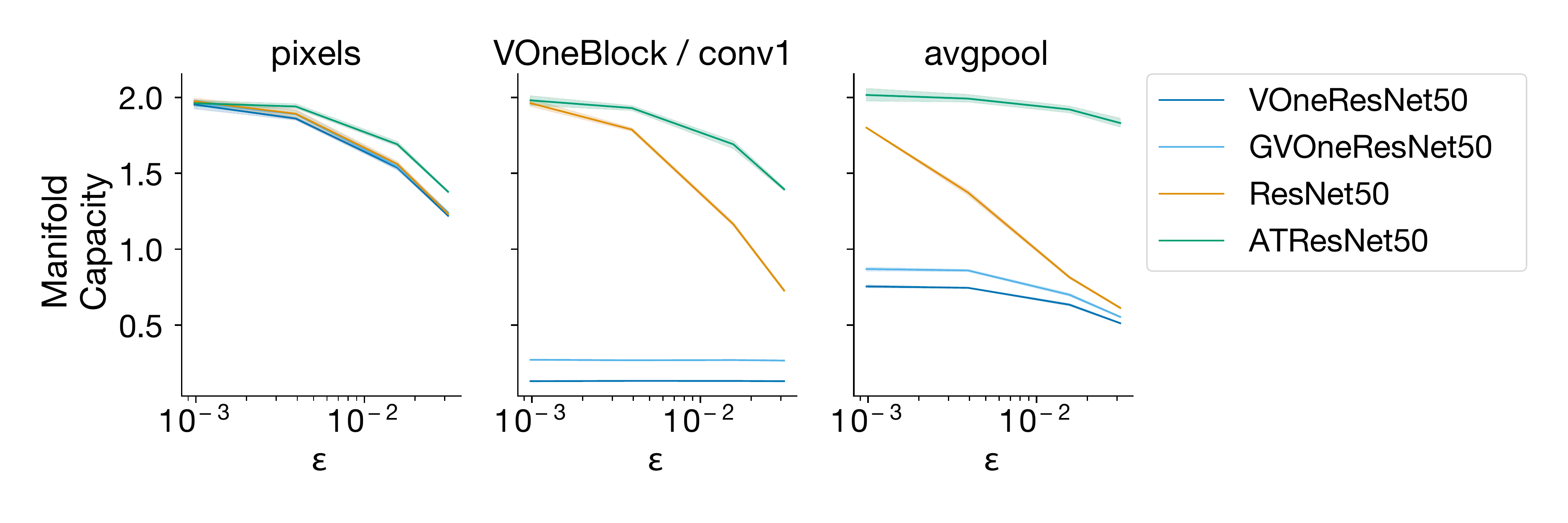}};
    \end{tikzpicture}
          \vspace{-4mm}
  \caption{\textbf{Unnormalized adversarial $\epsilon$-sized ball exemplar manifold capacity for pixels, VOneBlock / Conv1, and the final average pooling layer.} Figure 3C without normalization by the clean exemplar manifolds. Error bars are STD for 5 RP and MFTMA seeds.}
  \label{fig:unnormalized_cap_vs_eps}
\end{figure}

Providing additional context for Figure 3C, Figure \ref{fig:unnormalized_cap_vs_eps} shows the raw manifold capacity (not normalized by the capacity for clean exemplars) of VOneResNet50, GVOneResNet50, ResNet50, and ATResNet50. All networks and all layers considered are above the theoretical lower bound of 0.04 for capacity (given by $2/M$, where M is the number of example points in each manifold.)

When constructing Figure 3C, the values in \ref{fig:unnormalized_cap_vs_eps} were divided by the clean exemplar manifold capacity for each network (ie manifolds measured from unperturbed stimuli and containing variability only from the stochastic activations, if stochasticity was present in the model). Clean exemplar manifold capacity is set to the theoretical upper bound of 2 in all deterministic layers, which is equivalent to treating the manifold as a point. For VOneResNet50, clean exemplar manifold capacity is 0.13214 at the VOneBlock, and 0.90949 at the average pooling layer. For GVOneResNet50, clean exemplar manifold capacity is 0.27230 at the VOneBlock, and 1.01192 at the average pooling layer.

\subsection{Normalized exemplar capacity for additional models}\label{sec:additional networks}

\begin{figure}[b!]
  \centering
    \begin{tikzpicture}
    \pgftext{\includegraphics[width=0.9\columnwidth]{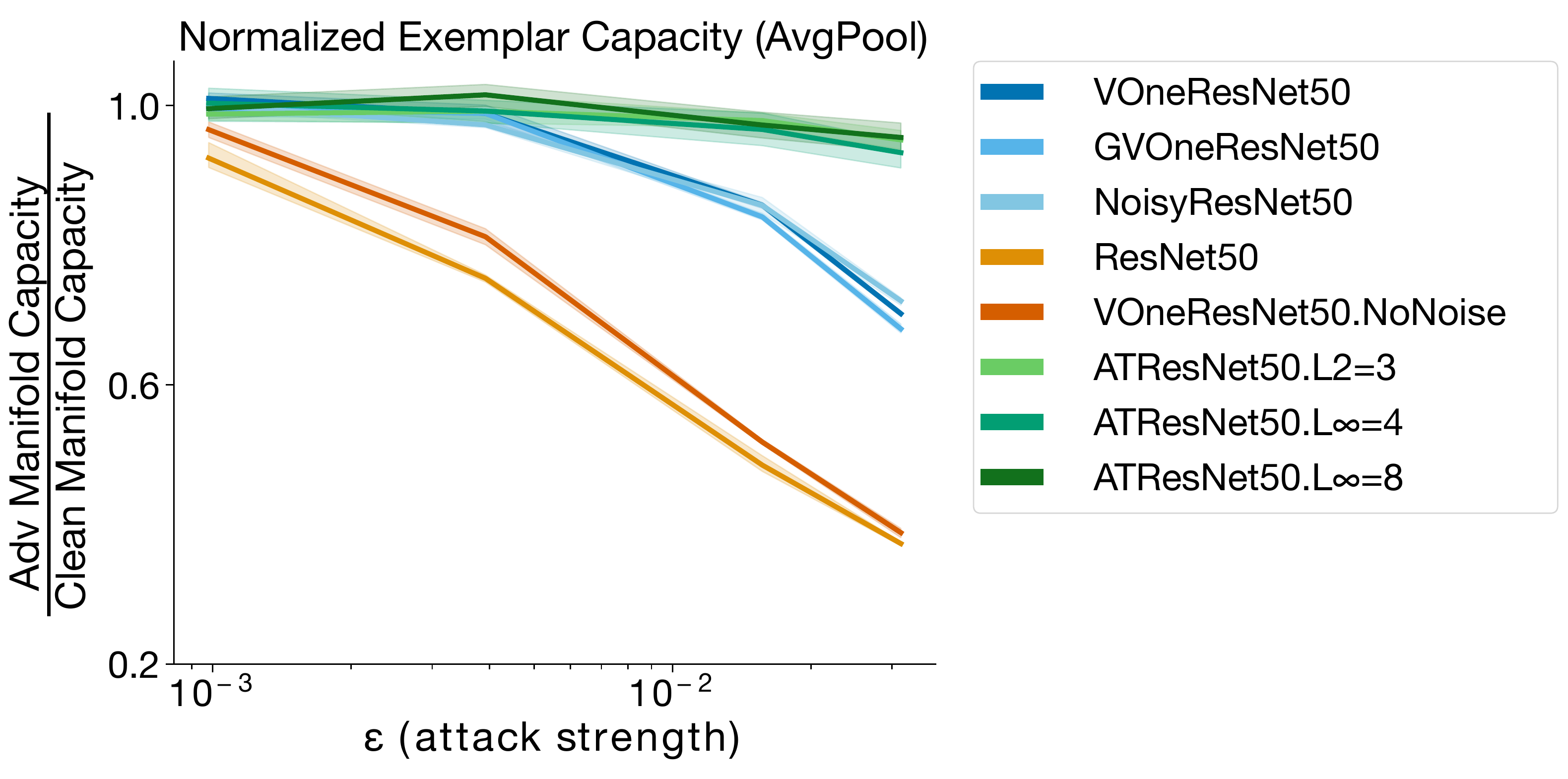}};
    \end{tikzpicture}
          \vspace{-4mm}
  \caption{\textbf{Normalized exemplar capacity for additional models.} relative exemplar manifold capacity (adversarial exemplar manifold capacity normalized by clean exemplar manifold capacity) is show for the average pool layer of adversarially trained models in green including ATResNet50.$L_2=3$ (light green), ATResNet50.$L_\infty=4$ (green), and ATResNet50.$L_\infty=4$ (dark green), stochastic models in blue VOneResNet50 (blue), GVOneResNet50 (light blue), and NoisyResNet50 (lighter blue), and non-stochastic, non-adversarially trained models in orange including ResNet50 (orange) and VOneResNet50.NoNoise (dark orange) show that stochasticity and adversarial training both lead to unique signatures of robustness across a range of networks.}
  \label{fig:additional_networks}
\end{figure}

Here, we include several additional ImageNet trained models to confirm the generality of our findings. To look at adversarially robust models beyond the ResNet50 adversarially trained with an $L_\infty$ norm of 4/255 (commonly ATResNet50, here ATResNet50.$L_\infty=4$) used in the main text, we investigate two additional adversarially trained ResNet50s, one with a stronger $L_\infty$ penalty of 8/255 (ATResNet50.$L_\infty=8$), and another with an $L_2$ penalty of 3.0 (ATResNet50.$L_2=3$)\footnote{Additional adversarially trained models were taken from \url{https://github.com/MadryLab/robustness} \cite{robustnesslib}}. To isolate the influences of stochasticity compared to the influence of the fixed VOne representation, we also include a ResNet50 with stochasticity added after the initial conv-relu-maxpool block (NoisyResNet50, see \cite{dapello2020simulating} for full details), and a VOneResNet50 without stochasticity during training or inference (VOneResNet50.NoNoise, see \cite{dapello2020simulating} for full details.) These are compared to ResNet50, VOneResNet50 and GVoneResNet50 from the main text. As shown in Figure \ref{fig:additional_networks}, the adversarial trained models in green all travel together, the noisy models in blue travel together, and the non-stochastic, non-adversarially trained models in orange also travel together, reflecting distinct geometries due to adversarial training and in models with stochastic activations.

\clearpage
\section{Auditory networks}\label{sec:auditory_networks}

\subsection{Model architecture and training details}
\label{sec:auditorytraining}
All auditory models (CochResNet50, ATCochResNet50, and StochCochResNet50) included all components of the standard ResNet50 architecture \cite{He2015resnet}. Rather than the first convolutional layer acting on an image, the first convolution is applied to the generated cochleagram representation.

The `cochleagram' representation is similar to a spectrogram, but with frequency resolution and compression tuned to approximate the human ear. Cochlear filters were constructed using the pycochleagram library (\url{https://github.com/mcdermottLab/pycochleagram}). The cochlear model consists of a filterbank of 211 bandpass filters with frequency response as the positive portion of a cosine function, spaced on an equivalent rectangular bandwidth (ERB) scale with low limit of 50Hz and high limit of 10kHz, including lowpass and highpass filters \cite{mcdermott2011sound, glasberg1990derivation}. Audio input to the networks was two seconds long sampled at 20kHz (40,000 samples). Passing audio through these filters results in audio subbands, and the envelope of the each is computed via the analytic amplitude of the Hilbert Transform. Envelopes are downsampled to 200Hz and passed through a compressive nonlinearity ($x^{0.3}$). The output of this yields a cochleagram representation of shape (211, 390), which served as the input to the standard ResNet50 architecture layers. The cochleagram operations were implemented in PyTorch, and all components of the cochleagram generation are differentiable, allowing adversarial examples to be generated directly on the waveform in an end-to-end manner.

In StochCochResNet50, a layer of additive Gaussian noise with a mean of zero was applied to the cochleagram representation before being passed to the first convolutional layer of the ResNet50 architecture. For auditory model analyses, we investigated the audio input to the network (waveform), the output of the cochlear model (cochleagram), the additive Gaussian noise stochastic layer (cochleagram + noise), the first conv-relu (conv1) of ResNet50, the first maxpool (maxpool) of ResNet50, the output of each residual block (block1, block2, block3, block4), and the average pooling layer (avgpool) that occurs before the logits of the ResNet50 architecture. To align layerwise plots, the "cochleagram (+ noise)" layer is a duplication of the "cochleagram" layer in deteministic models CochResNet50 and ATCochResNet50, while it is the "cochleagram + noise" for StochCochResNet50.

Auditory networks were trained in PyTorch 1.5.0 on the word recognition task from the Word-Speaker-Noise dataset introduced in \cite{feather2019metamers}. Audio sampling rate was 20kHz. Samples from the audioset dataset served as additive noise in the waveform, and were combined with the speech clips at uniformly selected signal-to-noise ratios of -10dB to 10dB. A random 2 second crop of the speech signal was extracted, always ensuring that the labeled word overlapped with the 1 second boundary of the signal, and a random 2 second crop of the audioset background was added to the speech signal. The combined audio was mean subtracted and normalized to a root-mean-square level of 0.1 before being processed by the cochlear model. Adversarial performance curves were evaluated on held out speech clips, randomly cropped and normalized the same as during training (but excluding the audioset augmentation).

Models were trained with a batch size of 256 on 8 Nvidia Tesla-V100 GPUs on the MIT BCS OpenMind computing cluster. Stochastic models and the standard network took approximately 18 hours to train, while the adversarially trained network took approximately 114 hours. Each model was trained with 150 epochs of the speech data (corresponding to 42 epochs of the audioset clips). Learning rate started at 0.1 and was divided by 10 after every 50 epochs, using the pytorch SGD optimizer with momentum 0.9, weight decay 0.0001, and a cross entropy loss between the word labels and model predictions. All training was performed with a modified version of the robustness library \cite{robustnesslib} with additions to handle auditory training.

The adversarial training parameters for ATCochResNet50 consisted of a $L_{\infty}$-norm bound of 0.001 on the waveform. Five attack steps were applied for the PGD attack during training, starting at a random location with a step size of 0.001/2.

\subsection{Adversarial attacks}
\label{sec:aud_adversarial_attacks}
The adversarial robustness of audio models (StochCochResNet50, CochResNet50, ATCochResNet50) to untargeted $L_\infty$ attacks (Figure 4 and Figure \ref{fig:PerformanceAllAudGaussEPS}) was evaluated using 32 PGD steps with step size of $\epsilon/5$. For the stochastic models (StochCochResNet50) model gradients were sampled eight times for each PGD iteration and averaged to obtain the step direction. Analysis of the adversarial robustness with this ensemble method was conducted with \cite{art2018}.

For adversarial $\epsilon$-sized ball exemplar manifolds in Figure 5, FGSM with a random starting location was used to measure 50 samples for each audio exemplar. The attacks were untargeted $L_\infty$ attacks with step size of $2 \epsilon$. For analyses of adversarial class manifolds in Figure 6A, a single adversarial example was generated for each audio sample in the class using 32 steps of PGD. An untargeted $L_\infty$ attack with a random starting location and step size of $\epsilon/5$ was used for constructing these adversarial class manifolds. Similar to VOneResNet50 and CIFAR-VOneNet, we did not average over multiple gradient samples when constructing the manifolds for the stochastic audio models, as for these experiments we were focused on generating samples from the adversarial manifolds rather than on evaluating the models defenses. Adversarial examples for both types of manifold experiments were obtained using \cite{robustnesslib}.

\subsection{Choice of Gaussian noise level for StochCochResNet50}
\begin{figure}[h!]
  \centering
    \begin{tikzpicture}
    \pgftext{\includegraphics[width=0.3\columnwidth]{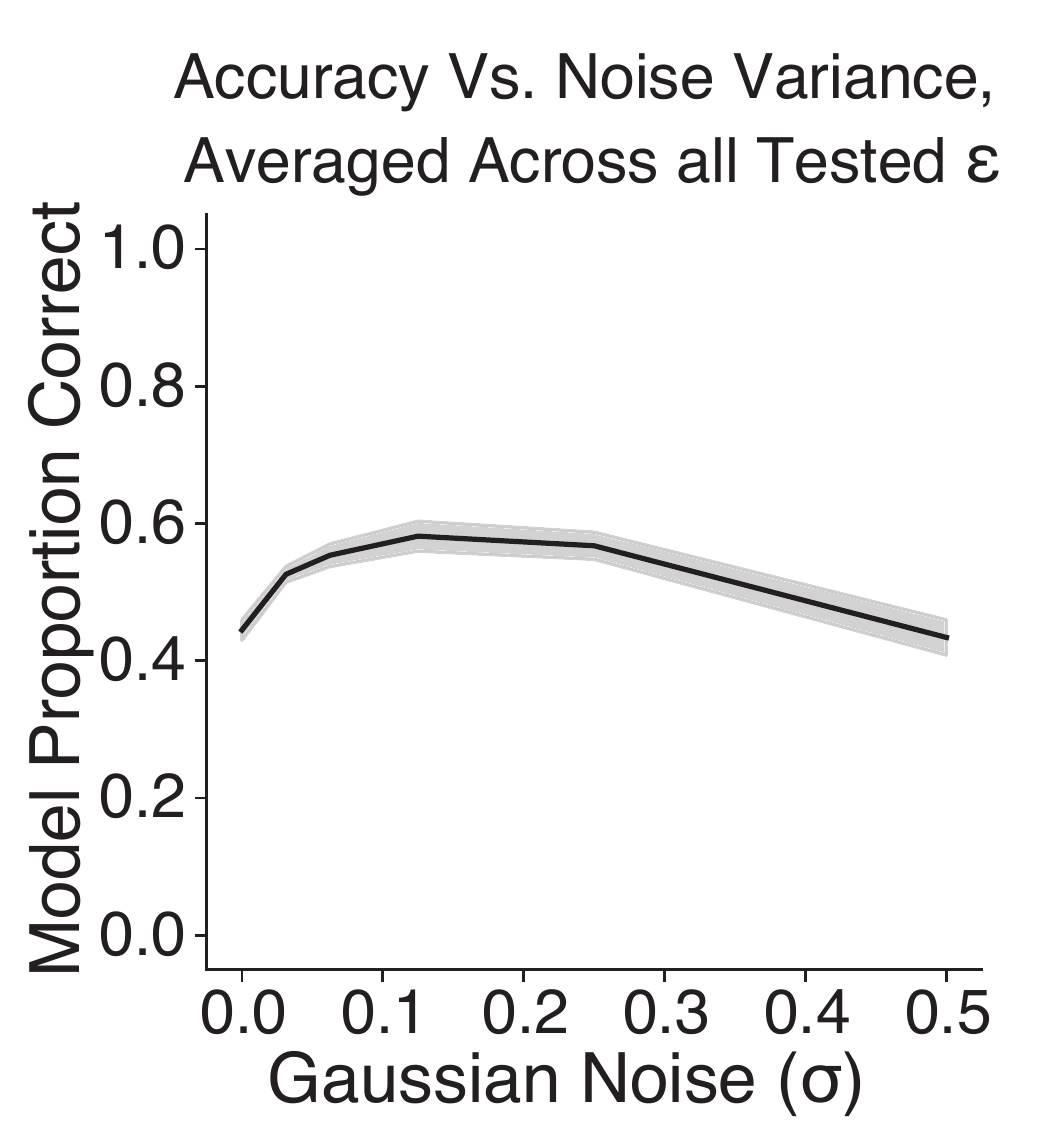}};
    \end{tikzpicture}
          \vspace{-4mm}
  \caption{\textbf{Adversarial performance of StochCochResNet50 models} Analysis of model performance averaged across all tested $L_\infty$ attack $\epsilon$ values compared to the level of Gaussian noise in the model, evaluated over 100 randomly chosen speech examples. Model performance peaks at $\sigma=0.125$. Error bars are STD across 5 sets of test stimuli.}
  \label{fig:PerformanceAllAudGaussEPS}
\end{figure}

The evaluation shown in Figure 4C showed that a Gaussian noise level of $\sigma=0.125$ yielded maximum adversarial robustness at $\epsilon=0.001$ to untargeted $L_\infty$ attacks. This level of Gaussian noise also yields best performance when averaged across all tested $\epsilon$ values (Figure \ref{fig:PerformanceAllAudGaussEPS}). Thus, Gaussian noise level of $\sigma=0.125$ was used for the presented experiments that made comparisons between StochCochResNet50, CochResNet50 and ATCochResNet50.

\begin{figure}[h!]
  \centering
    \begin{tikzpicture}
    \pgftext{\includegraphics[width=0.7\columnwidth]{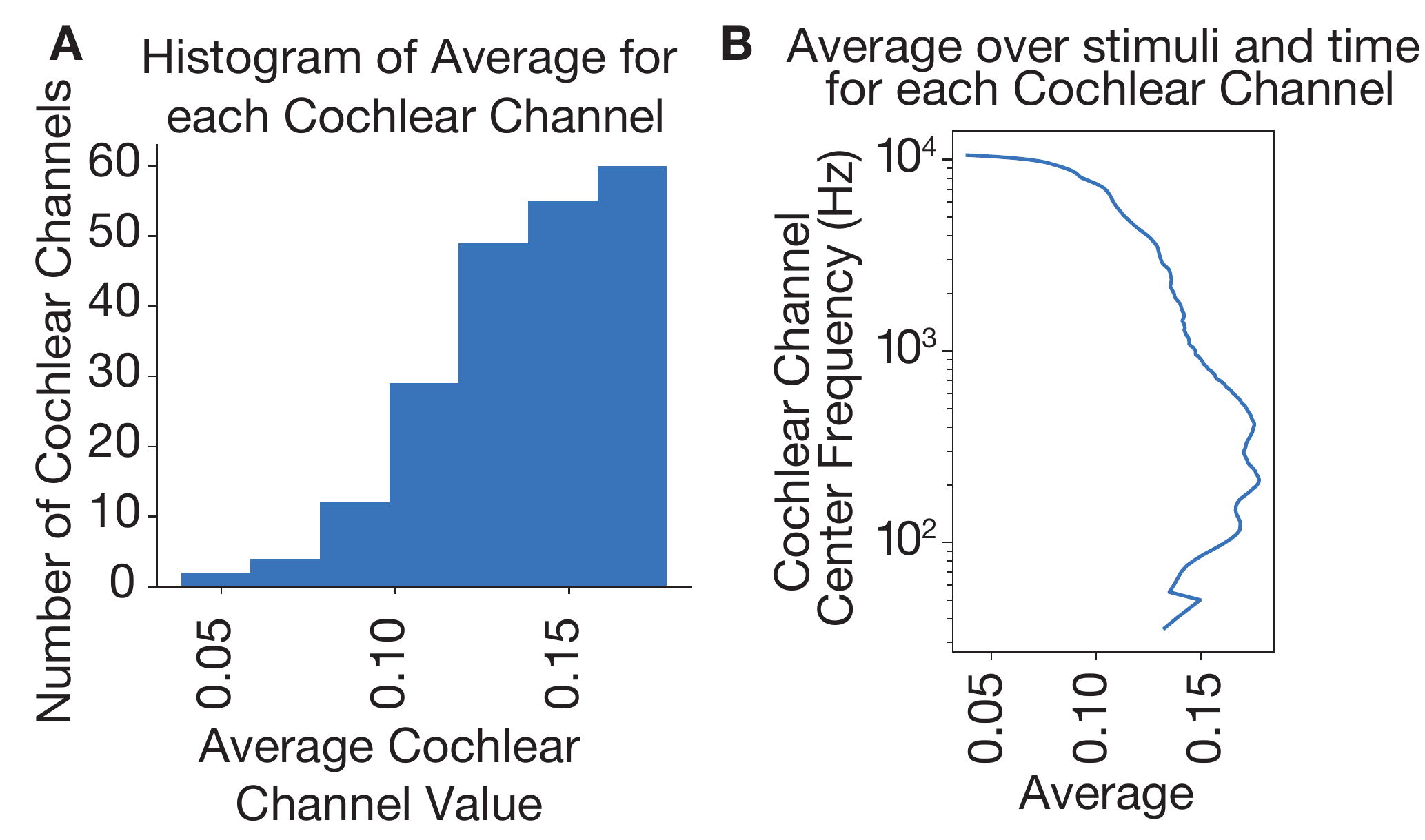}};
    \end{tikzpicture}
          \vspace{-4mm}
  \caption{\textbf{Average cochlear channel values}. (A) Histogram demonstrating the distribution of the average value across time in each cochleagram channel. (B) The time average of each cochleagram channel, ordered by frequency.}
  \label{fig:coch_channel_avg}
\end{figure}

We further quantified the the Signal-to-Noise-Ratio (SNR) of the stochastic cochleagrams for the selected Gaussian noise levels by (1) calculating the mean cochleagram across 20,000 examples from the training data (2) taking the mean across time and frequency (3) dividing this value by the standard deviation used for the Gaussian noise at the stochastic layer. These SNR values are reported in SM Table \ref{tab:SNRAudioNets}. Note that the best model shown in bold is close to the SNR ratio of 1 that was chosen for GVOneNet. A histogram of the averages for the cochlear channels and the ordered average for each cochlear channel is shown in Figure \ref{fig:coch_channel_avg} to demonstrate the distribution of average channel activations.

\begin{table}[h!]
  \centering
 \begin{tabular}{|p{0.35\columnwidth} | p{0.35\columnwidth}|}
 \hline
 Gaussian ($\sigma$) & StochCoch SNR \\ [0.5ex]
 \hline\hline
 0 & inf \\
 \hline
 0.03125 & 4.4433  \\
 \hline
0.0625 & 2.2217 \\
 \hline
 \textbf{0.125} & \textbf{1.1108} \\
 \hline
 0.25 & 0.5554 \\
 \hline
0.5 & 0.2777  \\
 \hline
\end{tabular}
\caption{SNR of the cochleagram + noise layer of the StochCochResNet50 architecture when changing the standard deviation ($\sigma$) of the additive Gaussian noise. Model with best performance across adversarial attacks is shown in bold. \label{tab:SNRAudioNets}}
\end{table}

\subsection{Characterizing adversarial robustness of StochCochResNet50}
\begin{figure}[b!]
  \centering
    \begin{tikzpicture}
    \pgftext{\includegraphics[width=0.95\columnwidth]{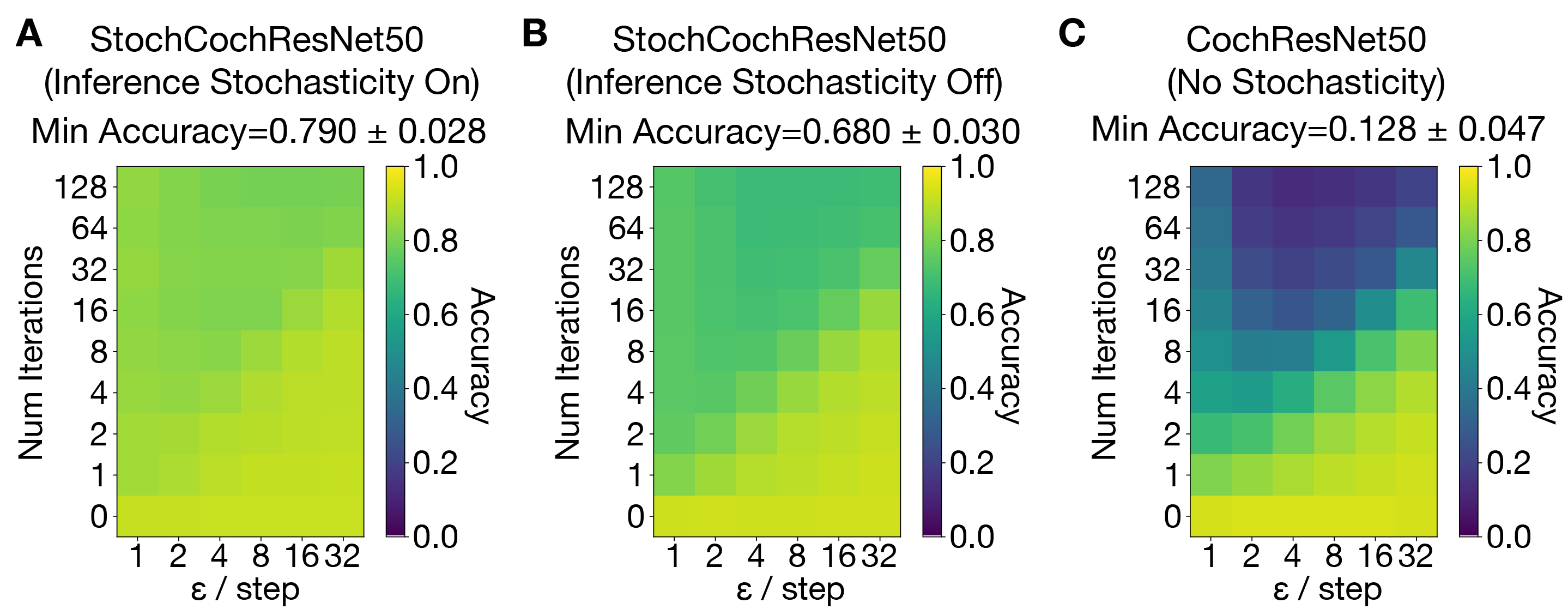}};
    \end{tikzpicture}
          \vspace{-4mm}
  \caption{\textbf{$L_{\infty}$ adversarial evaluation of Auditory Networks}. StochCochResNet50 is adversarially attacked with (A) and without (B) including stochasticity during adversarial generation and inference. CochResNet50 (C) is similarly adversarially attacked. The number of iterations and the size of the attack step is varied for each model, with $L_{\infty}$ attack strength of $\epsilon=0.001$. For StochCochResNet50 with inference, the gradients are averaged over 8 instantiations of the model. The worst accuracy across all attack possibilities is reported, and the STD is computed across 5 sets of test stimuli.}
  \label{fig:LinfSweepAttackIter}
\end{figure}

We further validated the adversarial robustness of the StochCochResNet50 with $\sigma=0.125$ to $L_{\infty}$ attacks by sweeping through different numbers of PGD iterations and step sizes (Figure \ref{fig:LinfSweepAttackIter}). The StochCochResNet50 with stochasticity during inference remained more robust than the StochCochResNet50 without stochasticity during inference, and both StochCochResNet50 evaluations were more robust than the CochResNet50.

As with our sanity checks on VOneResNet50 and GVOneResNet50, Figure \ref{fig:LinfSweepAttackIter} results demonstrate that going from random perturbations (zero iterations) to one PGD iteration increases the effect of the attack, and again from one to many iterations the effect of the attack is increased, indicating that the gradients are not broken and indeed contain information for computing adversarial perturbations.

We further tested the adversarial robustness of StochCochResNet50 to $L_2$ perturbations. Similar to the $L_\infty$ results in Figure 3D, the StochCochResNet50 with stochasticity during inference was more robust than the StochCochResNet50 without stochasticity during inference, however both were more robust to $L_2$ perturbations than the the CochResNet50 with no adversarial defenses. The network trained with $L_{\infty}$ adversarial perturbations (ATCochResNet50) was more robust than all networks. We performed a similar analysis to Figure \ref{fig:LinfSweepAttackIter}, using $L_2$ attacks. Once again, for StochCochResNet50 with inference stochasticity, we see that by increasing the number of iterations the performance decreases, indicating that the gradients for the stochastic networks maintain information to compute adversarial perturbations. For all evaluations of StochCochResNet50 with inference stochasticity, we found that an ensemble size greater than eight no longer improves effectiveness of the adversarial attacks.

\begin{figure}[h!]
  \centering
    \begin{tikzpicture}
    \pgftext{\includegraphics[width=0.9\columnwidth]{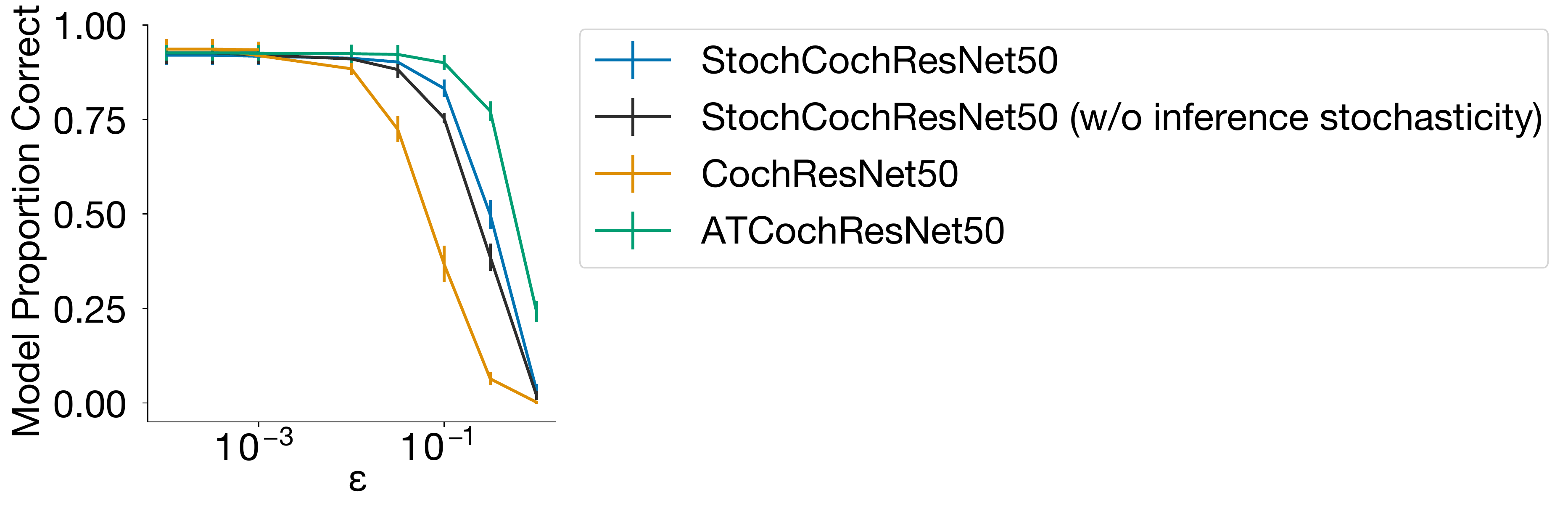}};
    \end{tikzpicture}
          \vspace{-4mm}
  \caption{\textbf{$L_2$ evaluation of Auditory Networks for different $\epsilon$ sizes}. Adversarial performance of auditory networks on $L_2$ adversaries for various $\epsilon$ sizes. Step size was set to $\frac{\epsilon}{8}$ and generated from 32 PGD iterations. For StochCochResNet50 with inference, gradients for each iteration were averaged over 8 instantiations of the model.}
  \label{fig:L2SweepEPS}
\end{figure}

\begin{figure}[h!]
  \centering
    \begin{tikzpicture}
    \pgftext{\includegraphics[width=0.95\columnwidth]{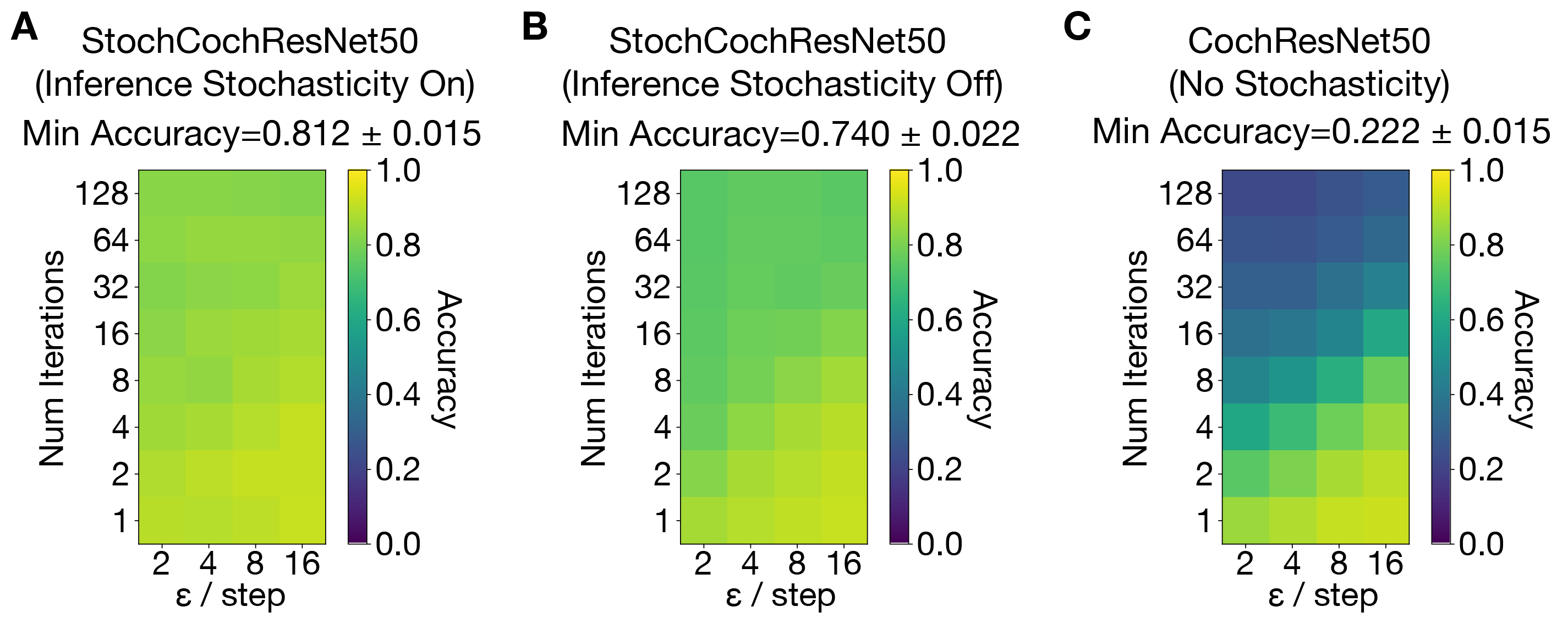}};
    \end{tikzpicture}
          \vspace{-4mm}
  \caption{\textbf{$L_{2}$ adversarial evaluation of Auditory Networks}. StochCochResNet50 is adversarially attacked with (A) and without (B) including stochasticity during adversarial generation and inference. CochResNet50 (C) is similarly adversarially attacked. The number of PGD iterations and the size of the attack step is varied for each model, with $L_{2}$ attack strength of $\epsilon=0.1$. For StochCochResNet50 with inference, the gradients are averaged over 8 instantiations of the model. The worst accuracy across all attack possibilities is reported, and the STD is computed across 5 sets of test stimuli.}
  \label{fig:L2SweepAttackIter}
\end{figure}

\subsection{Auditory Networks: Capacity for different adversarial strengths, unnormalized}\label{sec:unnormaudcap}
\begin{figure}[h!]
  \centering
    \begin{tikzpicture}
    \pgftext{\includegraphics[width=0.9\columnwidth]{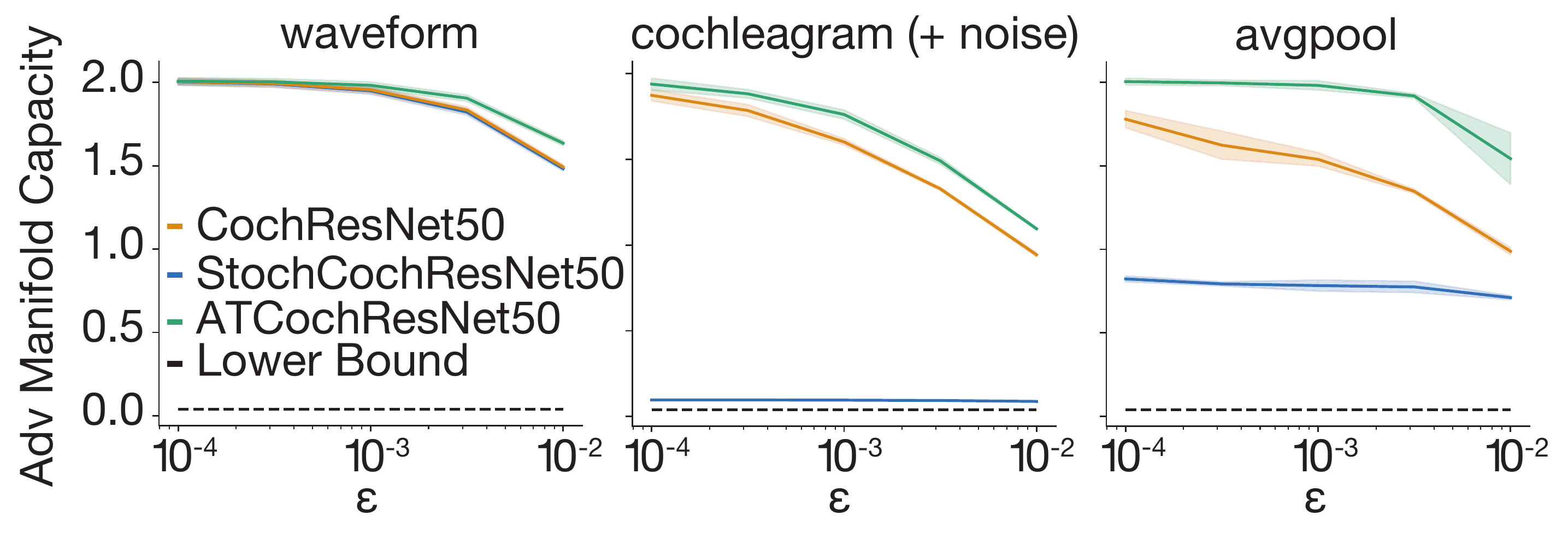}};
    \end{tikzpicture}
          \vspace{-4mm}
  \caption{\textbf{Auditory networks adversarial exemplar manifold capacity, unnormalized} Analysis of auditory networks unnormalized $\epsilon$-sized ball exemplar manifold capacity for waveform, cochleagram (+ noise), and the final average pooling layer. Dashed line represents the lower bound for capacity. Error bars are STD across 5 RP and MFTMA seeds.}
  \label{fig:unnormalized_cap_vs_eps_aud}
\end{figure}

Figure \ref{fig:unnormalized_cap_vs_eps_aud} shows the raw manifold capacity (not normalized by the capacity for clean exemplars) of StochCochResNet50, CochResNet50, and ATCochResNet50. All networks and all layers considered are above the theoretical lower bound of capacity ($2/M$, where $M$ is the number of points in each manifold, here $M=50$.) Clean exemplar manifold capacity used for normalization in Figure 5B is 2 for deterministic layers, 0.09690 at the stochastic cochleagram layer of StochCochResNet50, and 0.80927 in the StochCochResNet50 average pooling layer. As in the VOneNet analysis, clean exemplar manifold capacity is set to the theoretical upper bound of 2 in all deterministic layers.

\subsection{Auditory Networks: Opposing factors of stochasticity level}
\begin{figure}[h!]
  \centering
    \begin{tikzpicture}
    \pgftext{\includegraphics[width=\columnwidth]{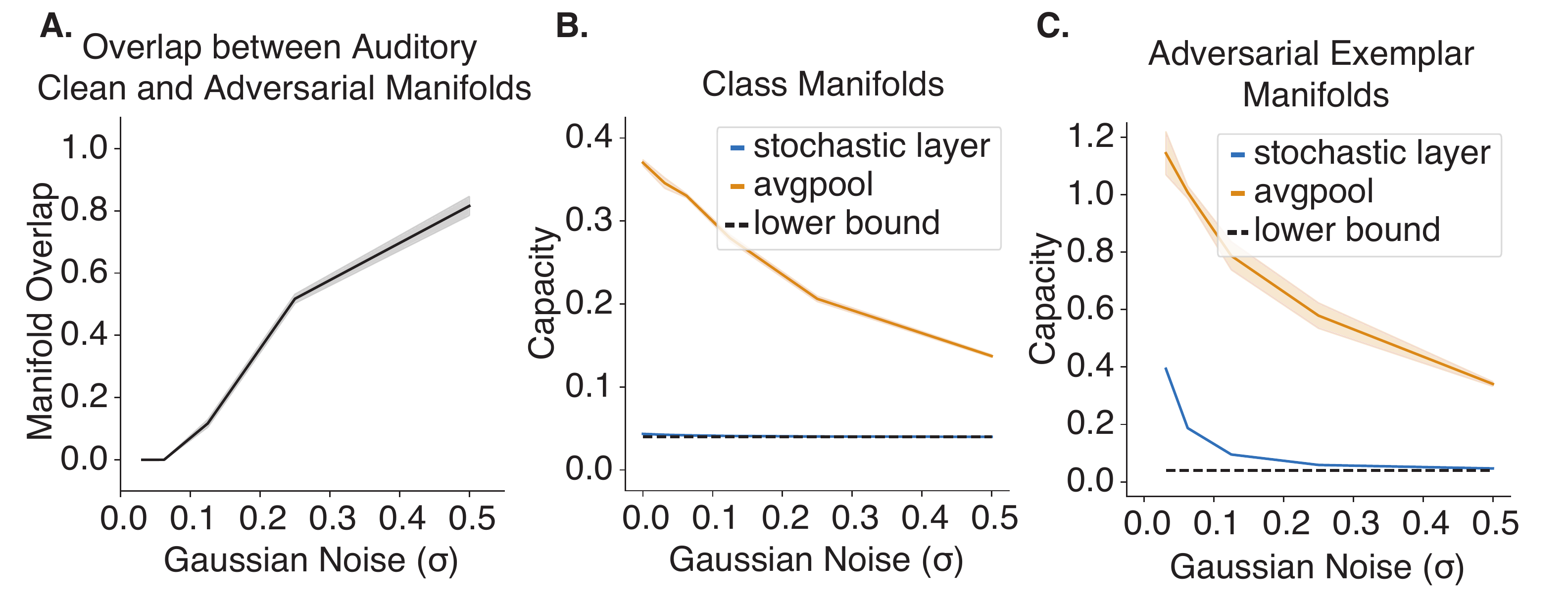}};
    \end{tikzpicture}
          \vspace{-4mm}
  \caption{\textbf{Auditory models demonstrate that stochastic representations induce opposing geometrical properties that determine a model's adversarial performance}. (\textbf{A}) SVM analysis of the overlap between the clean exemplar and adversarial exemplar manifolds measured at the stochastic "cochleagram + noise" layer of StochCochResNet50 trained and evaluated with varying levels of Gaussian noise. Error bars are STD across 10 sounds and 5 random seeds. (\textbf{B}) Class manifold capacity and (\textbf{C}) adversarial exemplar manifold capacity for the stochastic layer and the average pooling layer of a StochCochResNet50 trained and evaluated with varying levels of Gaussian noise. Error bars are STD across 4 random seeds.
}
  \label{figSM:AuditoryOpposingEffects}
\end{figure}

 We investigated whether the tradeoff seen in Figure 8 for the CIFAR network was also present for auditory networks. We examined manifolds created from samples within the $L_\infty$ $\epsilon=0.001$ ball, where we saw high variability in performance across networks with different $\sigma$. To measure the overlap of the clean exemplar and adversarial exemplar manifolds, we test whether an SVM can separate the exemplar noise manifolds from the adversarial perturbation exemplar manifolds generated from the same sound clip (Figure \ref{figSM:AuditoryOpposingEffects}A). Activations for 5000 samples of each exemplar are measured at the "cochleagram + noise" representation and downsampled to 2048 dimensions. Similar to the CIFAR results, as the noise increases the SVM error rate increases as the clean exemplar and adversarial exemplar manifolds become more entangled. The tradeoff with manifold capacity is revealed in the Class (Figure \ref{figSM:AuditoryOpposingEffects}B) and exemplar (Figure \ref{figSM:AuditoryOpposingEffects}C) manifold analyses. We analyze the class and adversarial exemplar manifolds from the stochastic layer and from the avgpool layer, constructing the manifolds as described in \ref{sec:aud_adversarial_attacks}. As the stochasticity in the network increases, the capacity decreases, until the noise outweighs the signal and the network cannot perform the task.

\begin{figure}[h!]
  \centering
    \begin{tikzpicture}
    \pgftext{\includegraphics[width=0.9\columnwidth]{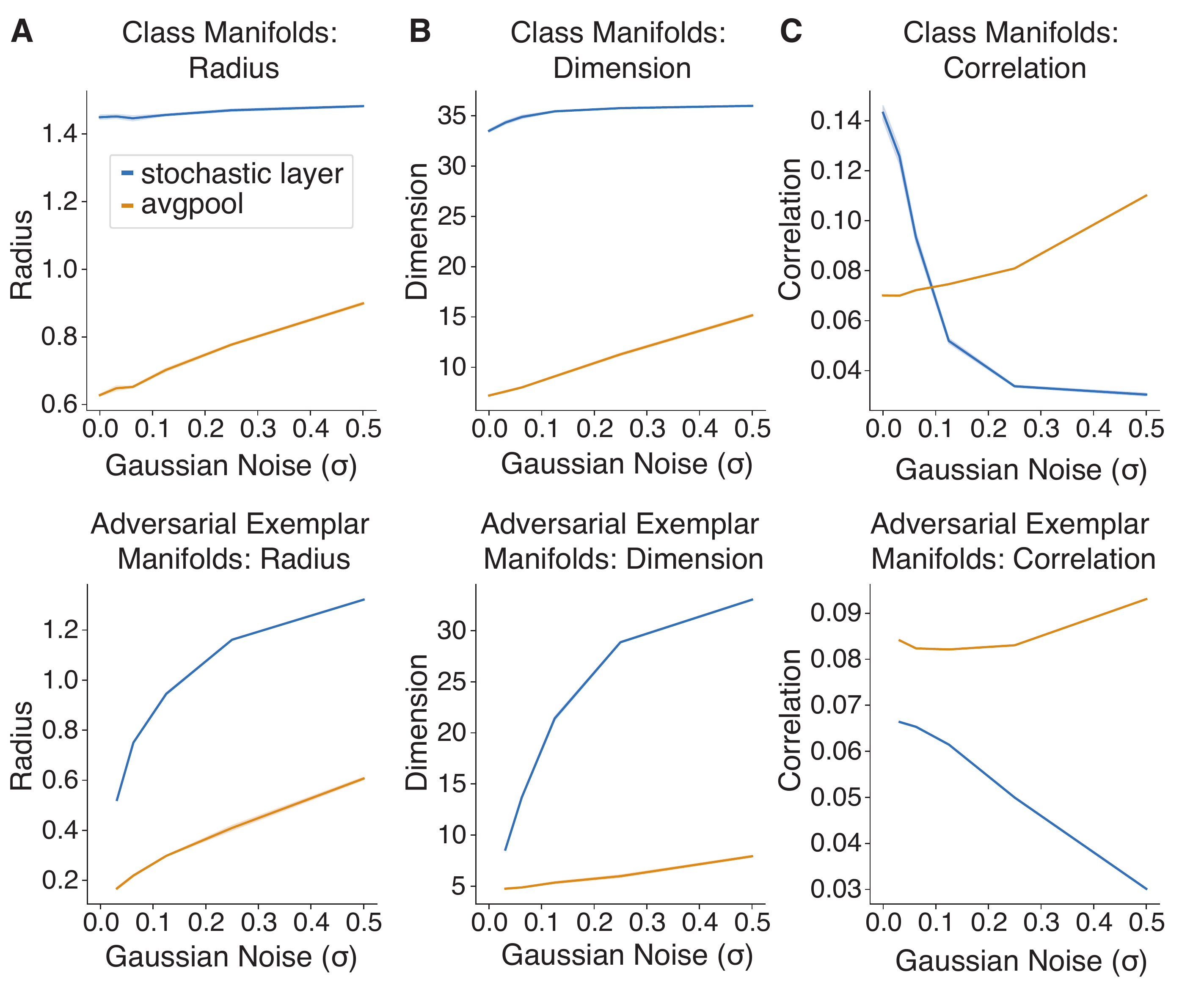}};
    \end{tikzpicture}
          \vspace{-4mm}
  \caption{\textbf{Additional geometric properties of StochCochResNet50 when varying level of Gaussian noise} \textbf{(A)} Manifold radius, \textbf{(B)} manifold dimension, (\textbf{C}), manifold center correlation of the Gaussian noise model with respect to the level of noise. Similar to the CIFAR results, as the level of noise increases, manifold radius and dimension also increases, while center correlation decreases at the early layer (after noise) and increases at the late layer (avgpool). Error bar is STD from 4 random seeds.}
  \label{fig:mftma_all_auditory_gaussian}
\end{figure}

Figure \ref{fig:mftma_all_auditory_gaussian} shows the MFTMA measured manifold radius, manifold dimension and center correlation for clean class manifolds and for adversarial exemplar manifolds across different Gaussian noise levels in StochCochResNet50 networks, corresponding to the capacity measures in Figure \ref{figSM:AuditoryOpposingEffects}B,C. Manifold radius and dimension increase as the noise increases for both types of manifolds, corresponding to an increased manifold width. The manifold center correlations decrease as noise increases for early layers of the network, but the manifolds become less correlated at late stages of the network with increasing noise.

\section{CIFAR10 vision networks}
\subsection{Model architecture and training details}


\begin{table}[h!]
\centering
    \begin{tabular}{| c | c | c |}
    \hline
    Layer Name & Description\\
    \hline
    pixels & Input $32 \times 32$ color image\\ [1ex]
    conv1 (prenoise) & VOneBlock 32 simple channels and 32 complex channels of shape 9 $\times$ 9, stride 1 \\ [1ex]
    conv1 (postnoise) & Gaussian noise layer \\ [1ex]
    block 1 & \texttt{conv2\_x} of ResNet18 \\ [1ex]
    block 2 & \texttt{conv3\_x} of ResNet18 \\ [1ex]
    block 3 & \texttt{conv4\_x} of ResNet18 \\ [1ex]
    block 4 & \texttt{conv5\_x} of ResNet18 \\ [1ex]
    avgpool & Average pool layer \\ [1ex]
    \hline
    \end{tabular}
\caption{CIFAR-VOneNet Architecture}
\label{tab:cifar_arch}
\end{table}

The CIFAR-VOneNet architecture is detailed in Table \ref{tab:cifar_arch}. For the VOneBlock, the architecture specification is based on the implementation of the VOneBlock in the VOnenet model from \cite{dapello2020simulating}. There are several modifications from the original VOneBlock. First, the parameters for the Gabor filters including the kernel size and the spatial frequency are scaled down to 50\% of their ImageNet size to better fit the small size of the CIFAR-10 dataset. Further, we use 32 simple and 32 complex channels instead of 256 simple and 256 complex channels. Finally, Gaussian noise, instead of Poisson noise, is added after the simple and complex cell non-linearities of the VOneBlock, for simplicity of tuning stochasticity levels. We note that the modification of the VOneBlock to fit CIFAR-10 inputs means that the original tuning to fit primate primary visual cortex data is disrupted, but we believe the general observations from our experiments on this version hold. We use \texttt{PyTorch 1.5} to train the models. The models are trained with a \texttt{SGD} optimizer with $120$ epochs and an initial learning rate of $0.1$. The learning rate is multiplied with a $\gamma$ factor of $0.1$ every $40$ epochs. The batch size is $128$ and the weight decay is $0.0005$. The models take approximately 3 hours to train with a NVIDIA Titan X GPU. All models were trained on the MIT BCS OpenMind computing cluster.

In Figure 8, we varied the stochasticity level by scaling the standard deviation of the Gaussian noise. Table \ref{tab:mnist_snr} provides the corresponding signal-to-noise (SNR) ratio to give more context on the magnitude of the stochasticity with respect to the strength of the feature activations. The signal strength is measured by taking the average activation at conv1 (prenoise) of 10000 images in the CIFAR test set. The signal-to-noise is calculated by dividing this average activation, which is 0.4215, by the Gaussian noise standard deviation. Note that the feature activation is quite sparse after the ReLU non-linearity activation at the end of the VOneBlock, so the mean feature activation may be smaller than the typical positive activation.

\begin{table}[ht!]
\centering
    \begin{tabular}{|c|c|}
    \hline
    Gaussian $\sigma$ & Activation SNR \\
    \hline
    0.01 & 42.152\\
    0.1 & 4.215\\
    1.0 & 0.422\\
    2.0 & 0.211\\
    4.0 & 0.105\\
    8.0 & 0.053\\
    10.0 & 0.026\\
    \hline
    \end{tabular}
\caption{Corresponding Signal to Noise Ratio with the standard deviation of the Gaussian noise}
\label{tab:mnist_snr}
\end{table}

\subsection{Adversarial attacks}
To evaluate the adversarial performance, adversarial images were generated by untargeted PGD with random starting locations within the $\epsilon$-ball. Specifically, the number of iterations was 64 and the step size was $\epsilon/32$. The model gradients used for each PGD step was the average gradient direction of 64 sample gradients to obtain reliable signal.

To generate the adversarial $\epsilon$-sized exemplar manifolds, untargeted FGSM with random starting locations within the $\epsilon$-ball was used. The number of iterations was 1 and the step size was $\epsilon$. The gradient used in FGSM was only sampled once, because in this experiment we focus on generating diverse samples for the adversarial manifolds rather than evaluating the model's adversarial robustness. All adversarial attacks for CIFAR models were performed using the adversarial robustness toolbox \cite{art2018}.

\subsection{Additional geometric properties for CIFAR-VOneNets}
In this section, we provide the additional MFTMA measurements including manifold dimensions, manifold radii, manifold center-center correlations, as well as the number of principal components needed to capture 90\% of the total representation variance for selected experiments detailing manifold capacity and width in Figure 8 of the main text.

\begin{figure}[h!]
  \centering
    \begin{tikzpicture}
    \pgftext{\includegraphics[width=\columnwidth]{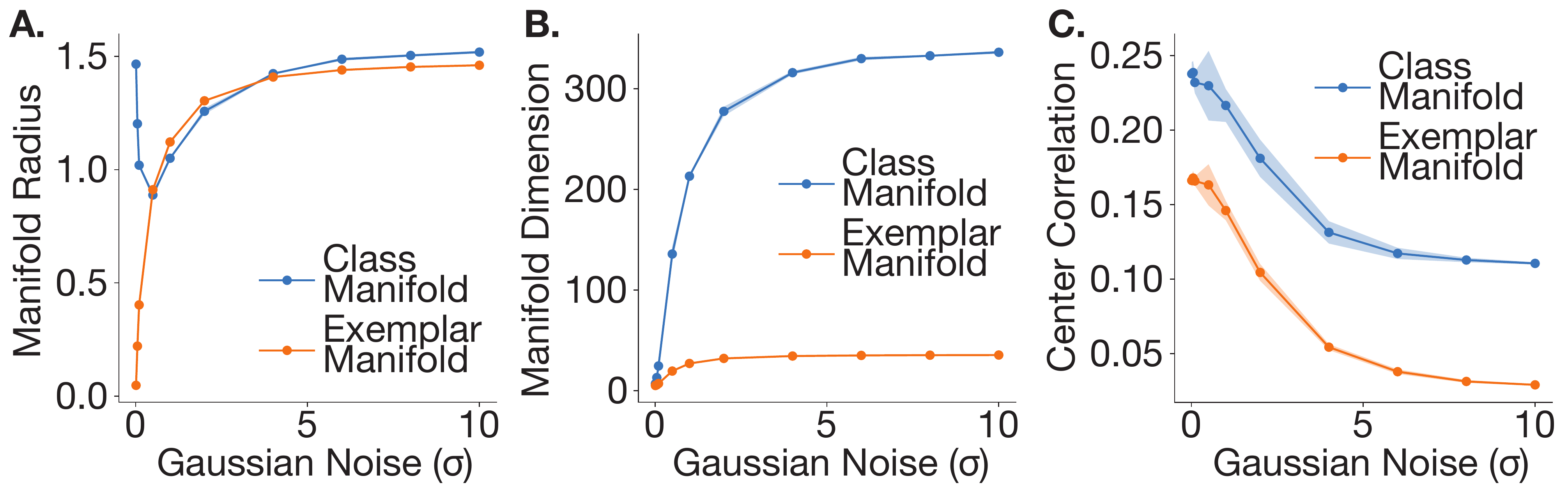}};
    \end{tikzpicture}
          \vspace{-4mm}
  \caption{\textbf{Additional geometric properties of CIFAR-VOneNet Gaussian noise model} \textbf{(A)} Manifold radius, \textbf{(B)} manifold dimension, (\textbf{C}) manifold center correlation of Gaussian noise model with respect to the level of noise, measured at the stochastic layer. As the level of noise increases, manifold radius and dimension also increases, while center correlation decreases. Error bar is STD from 6 RP and MFTMA seeds.}
  \label{fig:mnist_mftma}
\end{figure}

Figure \ref{fig:mnist_mftma} shows additional geometric properties (manifold radius, dimension and center correlation) for class manifolds and clean exemplar manifolds across different Gaussian noise levels in the CIFAR-VOneNet networks, measured at the stochastic conv1 (postnoise) layer.

\subsection{Replication of CIFAR-VOneNet model with Poisson noise}
\label{SM:CIFAR-poisson}

\begin{figure}[h!]
  \centering
    \begin{tikzpicture}
    \pgftext{\includegraphics[width=0.7\columnwidth]{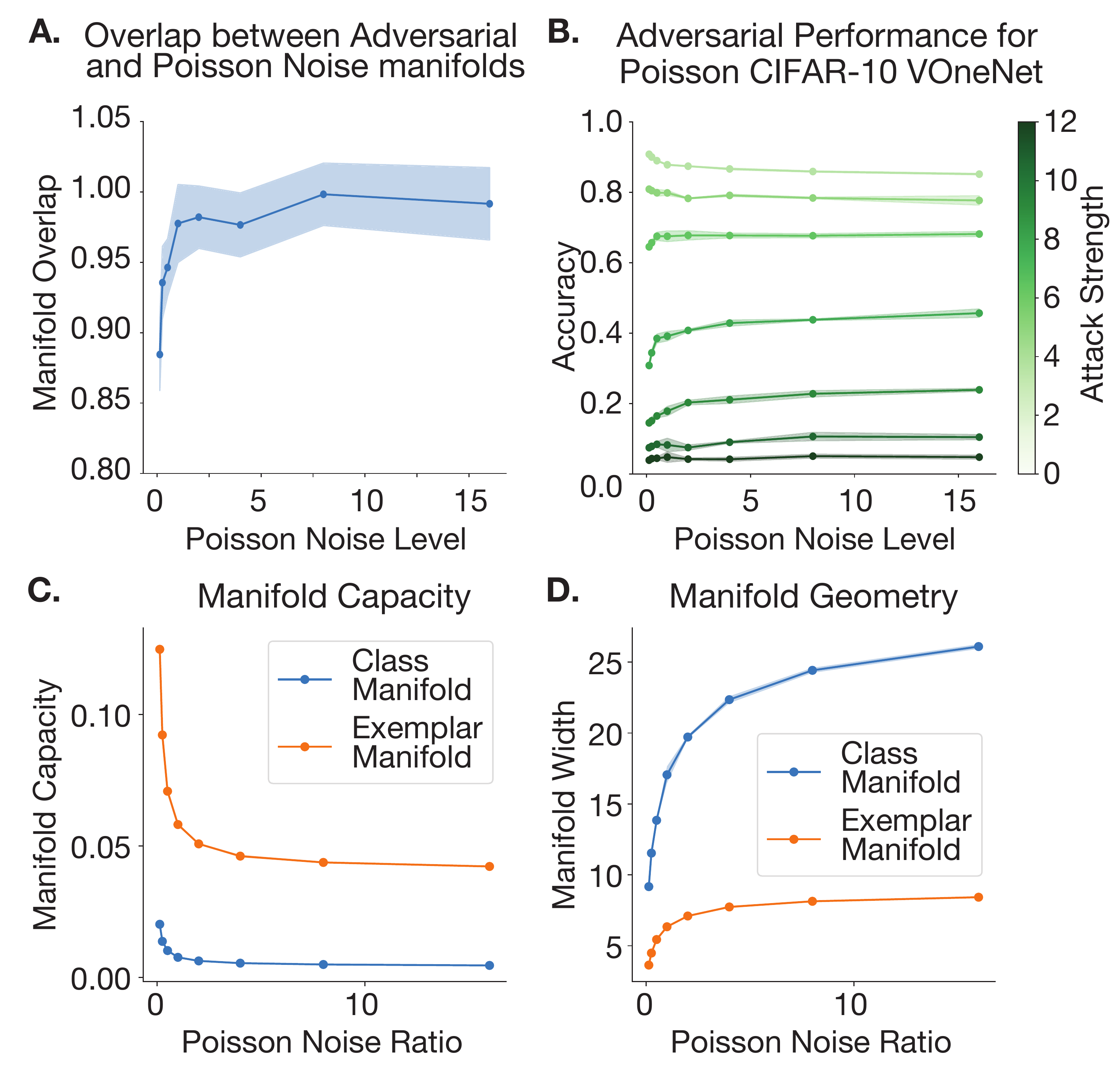}};
    \end{tikzpicture}
          \vspace{-4mm}
  \caption{\textbf{Poisson CIFAR-10 networks show opposing geometric effects of stochasticity} Replication of Figure 8 with Poisson CIFAR-10 networks. \textbf{(A)} A binary SVM is used to classify Poisson-like noise vs adversarial clouds from the same image, showing they become less linearly separable as the noise level increases, across multiple network layers. Error bar is STD across 20 images and 6 random seeds. \textbf{(B)} Adversarial performance across different attack strengths demonstrate the similar trade-off between clean accuracy and adversarial robustness. Error bar is STD across 6 random seeds. \textbf{(C)} Manifold Capacity decreases as the level of stochasticity increases. Error bar is STD across 6 random seeds. \textbf{(D)} Manifold Width increases as the level of stochasticity increases. Error bar is STD across 6 random seeds.}
  \label{fig:poisson_adv_perf}
\end{figure}

We replicate the overlap analysis presented in Figure 8 using a CIFAR VOneNet with Poisson noise rather than Gaussian noise (\ref{fig:poisson_adv_perf}). Similar to the Gaussian noise model, in the Poisson noise model, the noise is injected after the simple and complex cell non-linearities in the VOneBlock. We use the Poisson noise implementation from \cite{dapello2020simulating}, in which the Poisson noise is approximated by adding a Gaussian noise with mean zero and variance equal to the feature activation. With this formulation of Poisson noise, the variance-to-mean ratio is always 1. To investigate the effect of modifying the level of Poisson noise, we introduce the parameter noise ratio $r$, which represents the variance-to-mean ratio of the injected noise. For example, for the Poisson noise model with $r=2$, the added noise is Gaussian with mean zero and variance equal to twice the feature activation.

\begin{figure}[h!]
  \centering
    \begin{tikzpicture}
    \pgftext{\includegraphics[width=\columnwidth]{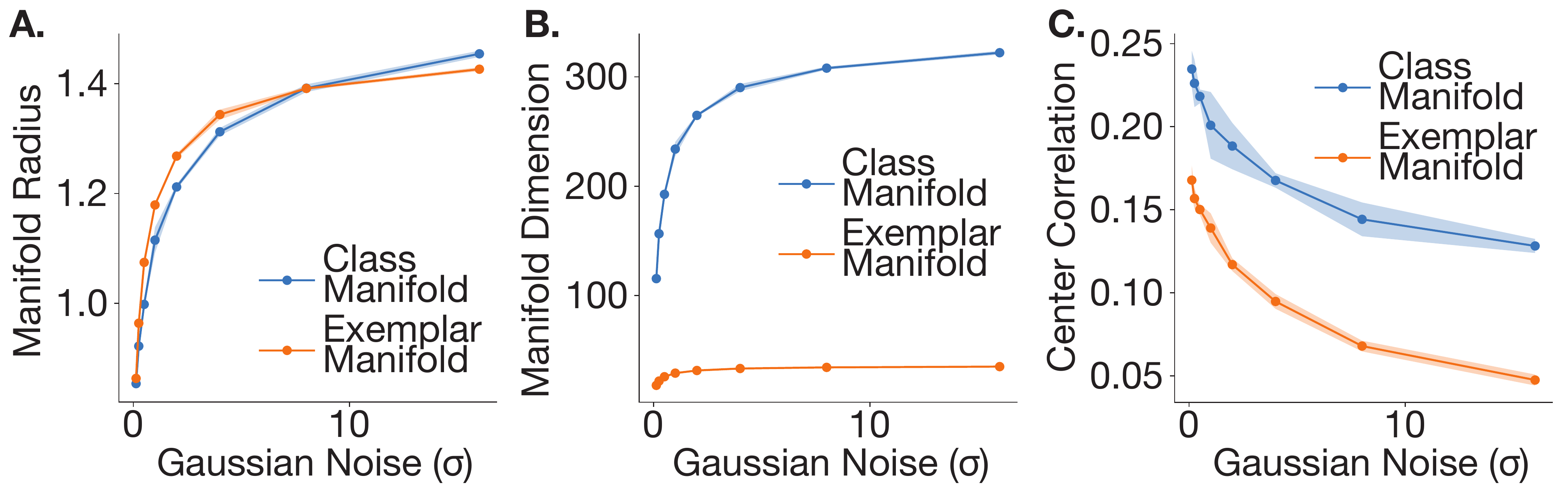}};
    \end{tikzpicture}
          \vspace{-4mm}
  \caption{\textbf{Additional geometric properties of Poisson-like noise model} \textbf{(A)} Manifold radius, \textbf{(B)} manifold dimension and \textbf{(C)} manifold center correlation of Poisson-like noise model with respect to the level of stochasticity. Manifold radius and dimension increases as the level of stochasticity increases while center correlation decreases. Error bar is STD from 6 RP and MFTMA seeds.}
  \label{fig:poisson_geometry}
\end{figure}

Similar to the Gaussian noise model, as the level of stochasticity increases, the stochastic exemplar adversarial and clean manifolds also become more overlapped (Figure \ref{fig:poisson_adv_perf}A). Figure \ref{fig:poisson_adv_perf}B shows the adversarial performance of the Poisson noise model with varying noise levels at multiple different attack strengths. Note that the model is trained and tested with the same noise levels. Compared to the Gaussian noise models, the performance of the Poisson noise model is less dependent on the noise level. Across various attack strengths, adversarial performance increases at low noise ratios and then plateaus, not degraded as in the case of Gaussian noise model, at high noise ratios. This phenomenon can be related to the property that Poisson noise has variance relative to individual unit activation, while Gaussian noise has an absolute variance for all units. Figure \ref{fig:poisson_adv_perf}C and Figure \ref{fig:poisson_adv_perf}D show MFTMA analysis for class manifolds and exemplar manifolds for the Poisson noise model with no adversarial perturbations, demonstrating that as the level of stochasticity increases, both the class and exemplar noise manifolds become more linearly entangled. Manifold radius, dimension, and center correlation are similarly shown in Figure \ref{fig:poisson_geometry}. Similar to Gaussian Noise, we see a general tradeoff between the overlapping adversarial and noise manifolds and the manifold capacity. However, compared to Gaussian noise, Poisson noise introduces different variances across individual units within the neural population, therefore introducing distinct population geometries. The nature of how different types of noise statistics affect robust perception through the lens of neural population geometry is an important direction for future studies.

\section{Replication of adversarial exemplar manifold results with random perturbations}\label{sec:randomepsballs}
Our exemplar manifold results detailed in the main text focused on investigating the geometry of activations measured from manifolds created by FGSM adversarial examples. One possible confound when defining manifolds based on adversarial examples is that different stimuli are used for each network (for instance, there may be a bias when generating the adversarial exemplar manifolds, resulting in some models having more similar examples within each exemplar manifold). As a control for this, we ran the the same exemplar manifold analyses using random points measured on the shell of the $L_{\infty}$ ball, presenting the same random points to each network. The main results held across these analyses, as documented in the following section. This highlights that the differences in geometry of the networks are measurable with generic stimuli and not a bi-product of the specific set of adversarial examples or attack methods used for the analysis.

\subsection{VOneNets}
\begin{figure}[h!]
  \centering
    \begin{tikzpicture}
    \pgftext{\includegraphics[width=0.7\columnwidth]{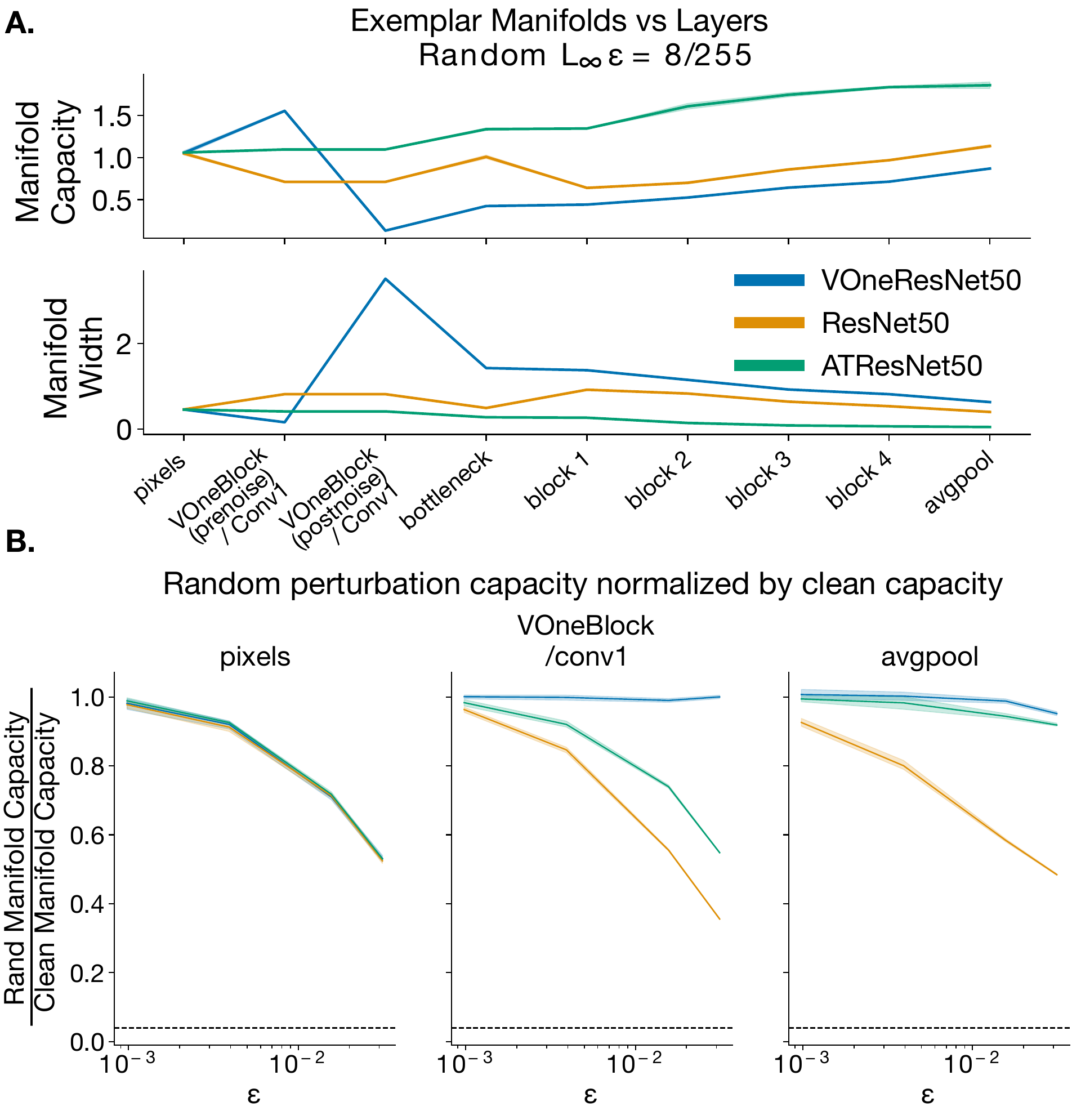}};
    \end{tikzpicture}
          \vspace{-4mm}
  \caption{\textbf{Random $\epsilon$-sized perturbations instead of FGSM-based perturbations for exemplar manifold analysis yields largely the same result} (\textbf{A}) Equivalent to Figure 3B but random corner perturbations are used instead of FGSM. (\textbf{B}) Equivalent to Figure 3C but random corner perturbations are used instead of FGSM. Error bars for all plots represent STD over 5 RP and MFTMA seeds.}
  \label{fig:vonenet_rand_eps_ball}
\end{figure}

Figure \ref{fig:vonenet_rand_eps_ball} replicates experiments from Figures 3B and 3C, except that all networks now receive the exact same perturbations drawn from random corners of the $L_\infty$ $\epsilon=8/255$ sized ball around an image ($\tilde{x} = x + \mathsf{sign}(\mathcal{N}(0,1)\epsilon$), demonstrating that while random perturbations cause generally less of an increase in capacity than FGSM-based perturbations, largely the trends are the same.

\subsection{Auditory Network}
\begin{figure}[h!]
  \centering
    \begin{tikzpicture}
    \pgftext{\includegraphics[width=0.7\columnwidth]{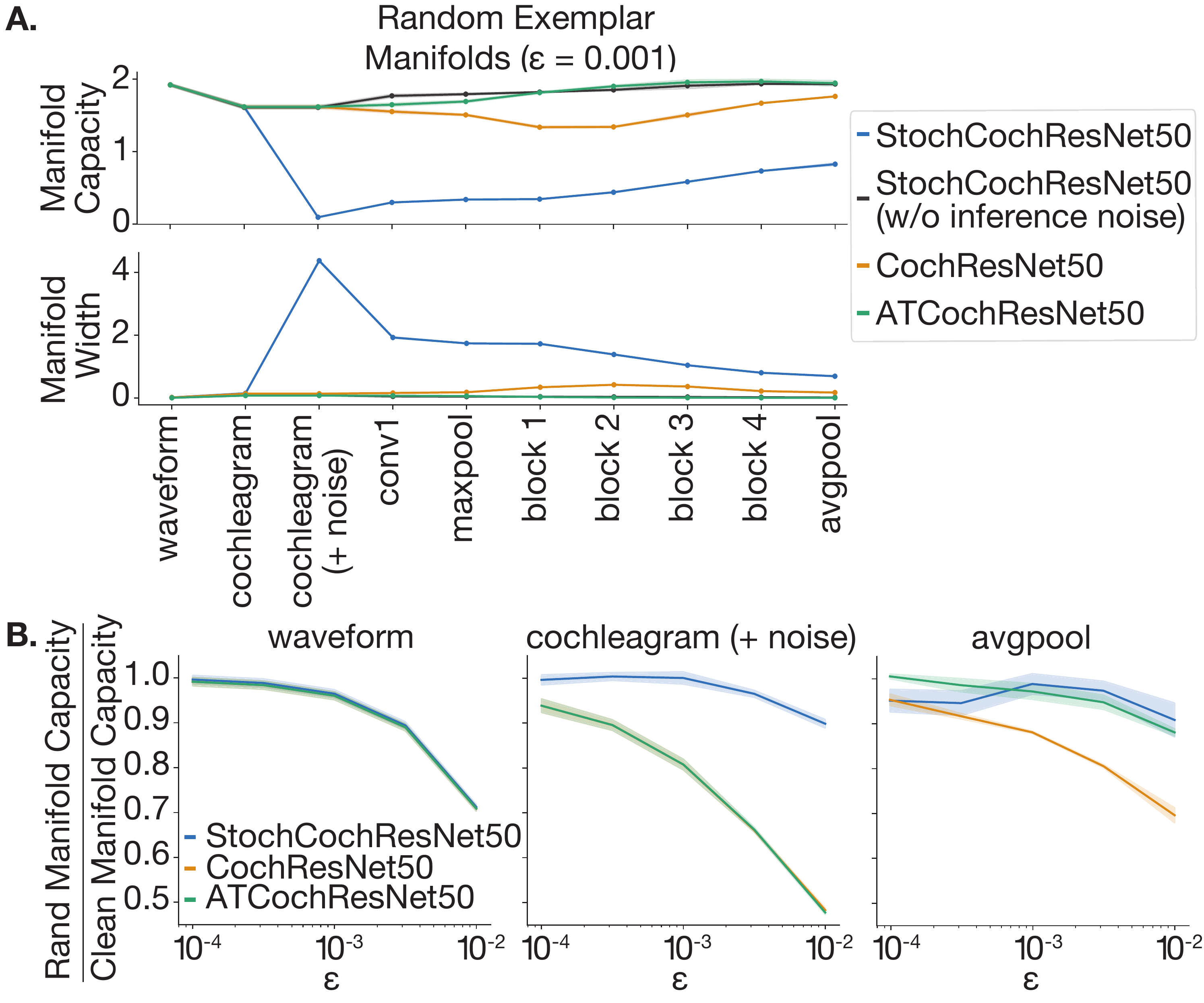}};
    \end{tikzpicture}
          \vspace{-4mm}
  \caption{\textbf{Random exemplar manifold geometry for auditory networks.} Replication of Figure 5 using random points from the shell of the $L_{\infty}$ ball at $\epsilon = 0.001$ to construct the exemplar manifold. (\textbf{A}) Capacity of random exemplar manifolds for layers of auditory models. For networks without stochasticity, the stochastic layer representation is equal to the cochleagram representation. Error bars are STD across 5 random projection seeds. (\textbf{B}) Random exemplar capacity normalized by the clean manifold capacity as a function of perturbation size. Error bars are STD across 5 random projection and MFTMA seeds. Data for StochCochResNet50, CochResNet50, and ATCochResNet50 is overlapping at waveform layer (as the same exemplar manifolds are tested for all networks) and data for CochResNet50 and ATCochResNet50 is overlapping at cochleagram layer, as this is also deterministic.}
  \label{fig:6e_randomepsballs}
\end{figure}

Auditory results conducted with adversarial exemplar manifolds (Figure 5) were replicated using random exemplar manifolds. Similar to the results with $L_{\infty}$ adversarial exemplar manifolds in Figure 5, we analyzed random exemplar manifolds constructed by taking points from the shell of the $L_{\infty}$ ball at $\epsilon=10^{-3}$ (Figure \ref{fig:6e_randomepsballs}A). In the average pooling layer, we see that the capacity for random exemplar manifolds is similarly high for ATCochResNet50 and StochCochResNet50 without inference stochasticity, and that the capacity of StochCochResNet50 with inference stochasticity is lower than CochResNet50. We saw similar replications for the normalized capacity as a function of the $\epsilon$ (Figure \ref{fig:6e_randomepsballs}B). Taken together with the findings from VOneNet, these results suggest that the measured geometric differences between networks cannot be due to differences in the manifold sampling and instead reflect differences in the internal transformations.

\clearpage
\section{Variability of manifold widths and random projections}

\begin{figure}[h!]
  \centering
    \begin{tikzpicture}
    \pgftext{\includegraphics[width=0.65\columnwidth]{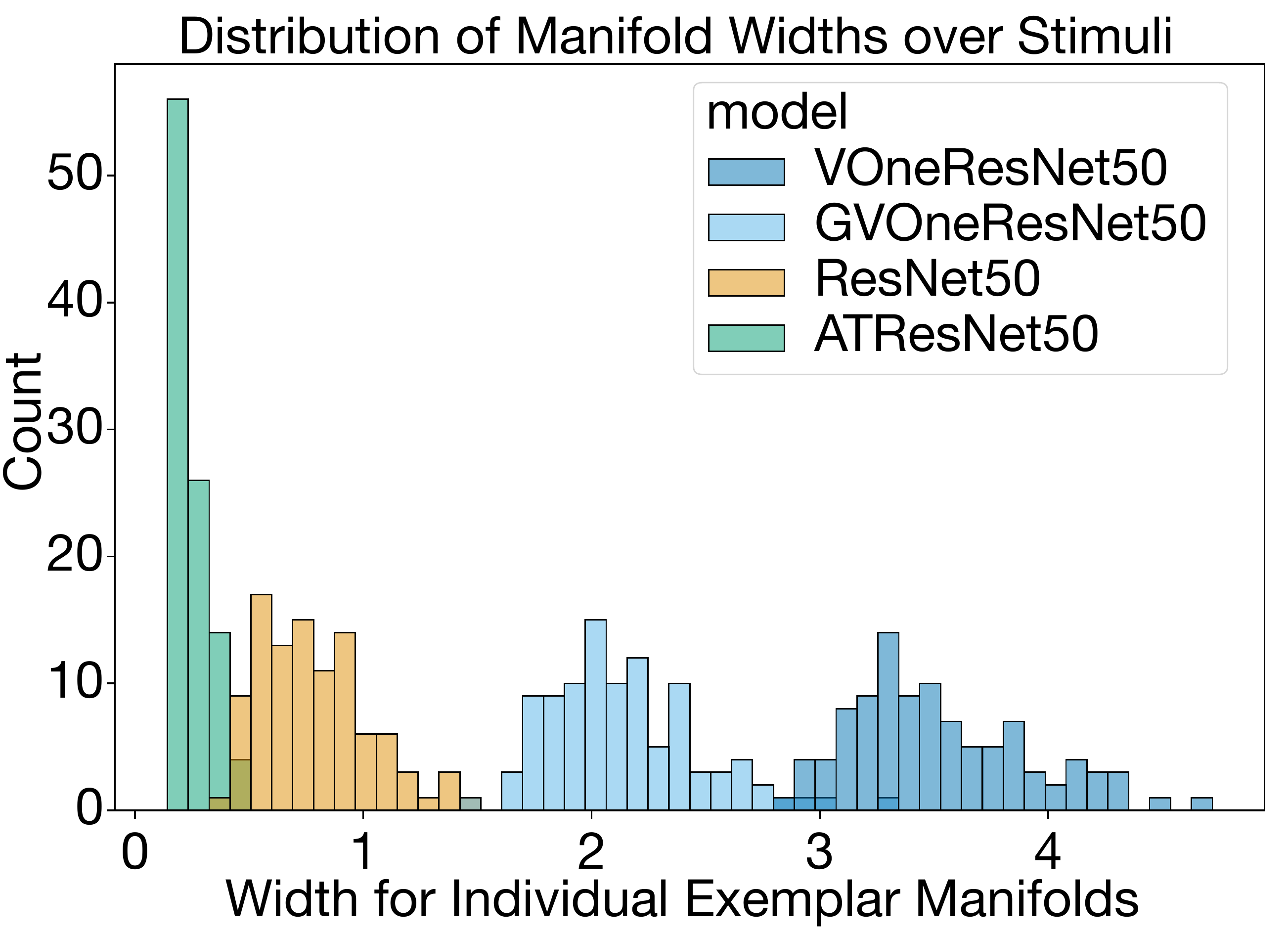}};
    \end{tikzpicture}
          \vspace{-4mm}
  \caption{\textbf{Raw distribution of exemplar manifold widths} The raw distribution of exemplar manifold widths for VOneResNet50, GVOneResNet50, ResNet50, and ATResNet50 at the VOneBlock (post noise) / conv1 layer show that manifold width for each network is reasonably well separated.}
  \label{fig:manifold_width_distribution}
\end{figure}

\begin{figure}[h!]
  \centering
    \begin{tikzpicture}
    \pgftext{\includegraphics[width=0.8\columnwidth]{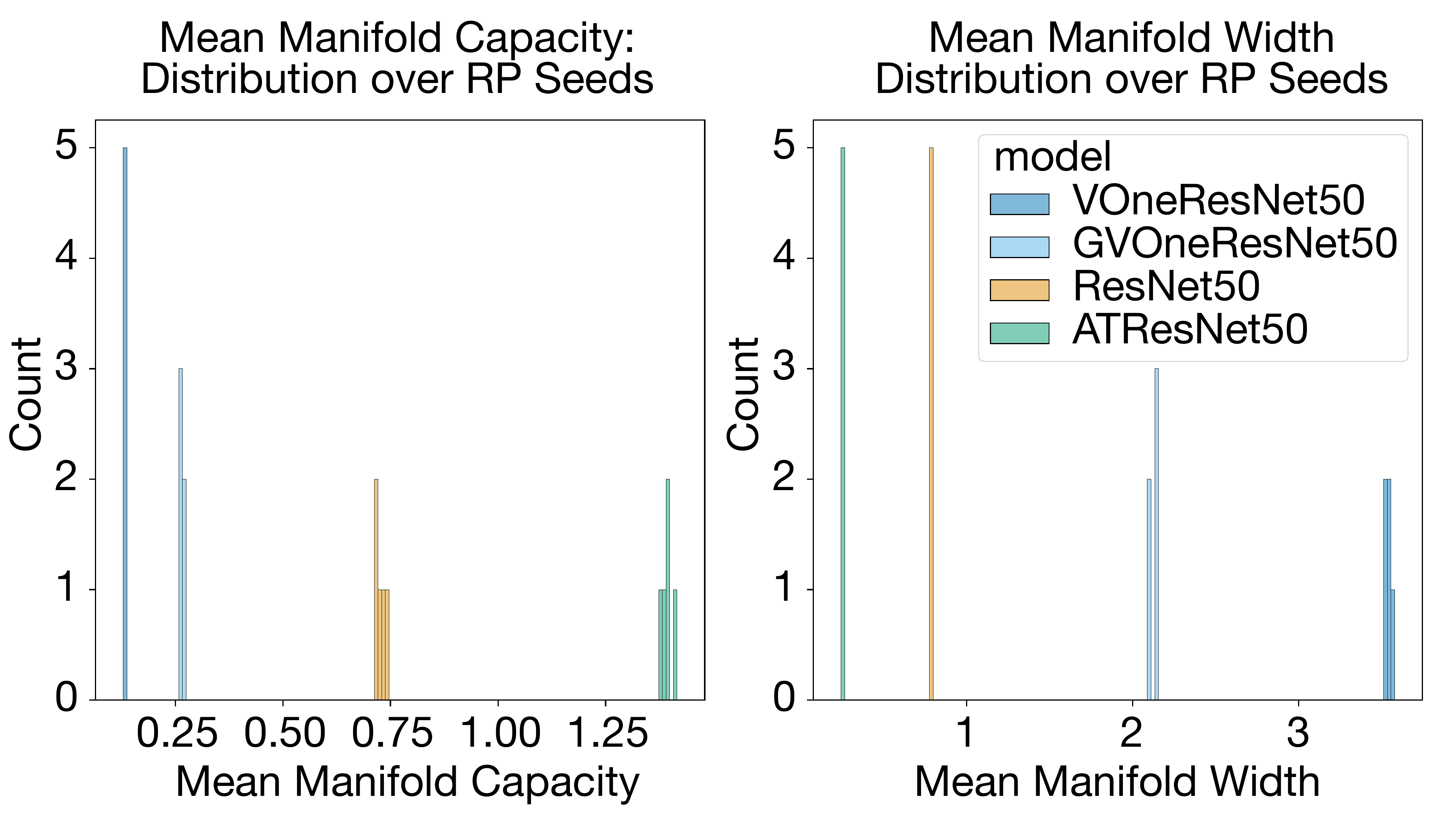}};
    \end{tikzpicture}
          \vspace{-4mm}
  \caption{\textbf{Raw distribution of RP and MFTMA seeds for exemplar manifold capacity and width} The raw distribution exemplar manifold capacity and width for 5 RP and MFTMA seeds for VOneResNet50, GVOneResNet50, ResNet50, and ATResNet50 at the VOneBlock (post noise) / conv1 layer show that variability from these sources is low and each network is well separated.}
  \label{fig:RP_distribution}
\end{figure}

In our main paper, the error bars for class and exemplar manifold analysis represent the STD of different random projection (RP) seeds and MFTMA seeds. In particular, this means that we have already averaged out the variability across different manifold widths before computing the STD error bars. Note, unlike manifold width, the full manifold capacity measure is a systems level property, and to be computed requires averaging over the manifold geometric properties. Thus, we chose to include error bars for width that similarly demonstrate a systems level property. To give a sense of the general distribution of exemplar manifold widths, in Figure \ref{fig:manifold_width_distribution} we show the raw distributions of exemplar manifold widths for VOneResNet50, GVOneResNet50, ResNet50, and ATResNet50 at the VOneBlock (post noise) / conv1 layer. We note that the distributions are well separated, and easily differentiable. In addition, we include Figure \ref{fig:RP_distribution}, which plots the raw distributions after RP and MFTMA seeds for the same layer and exemplar stimuli. Variability of RP and MFTMA seeds is quite low and again networks easily separable.

\section{Pairwise Distribution Analysis}
\begin{figure}[h!]
  \centering
    \begin{tikzpicture}
    \pgftext{\includegraphics[width=0.9\columnwidth]{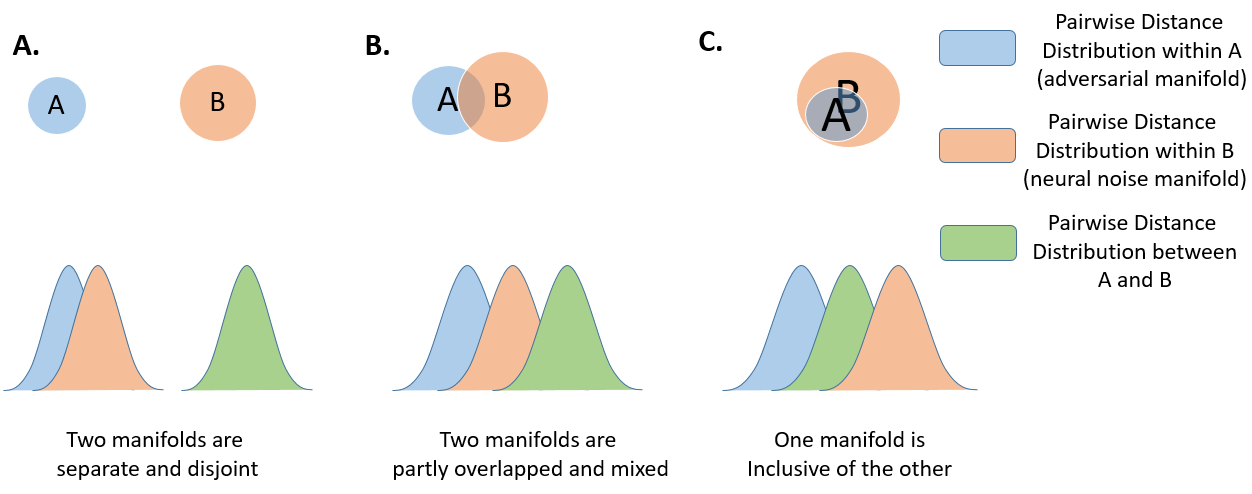}};
    \end{tikzpicture}
          \vspace{-4mm}
  \caption{\textbf{Illustration of pairwise distance analysis} The illustration shows the relative position of two clouds in the feature space and its corresponding pairwise distance distribution in three different scenarios: \textbf{(A)} two clouds are disjoint, \textbf{(B)} two clouds are partly overlapped and \textbf{(C)} one cloud is inclusive of the other. The more overlap between the between-manifold distribution and one of the within-manifold pairwise distance distributions, the more that these two clouds are overlapped in the feature space.}
  \label{fig:pairwise_illustration}
\end{figure}

\begin{figure}[h!]
  \centering
    \begin{tikzpicture}
    \pgftext{\includegraphics[width=0.9\columnwidth]{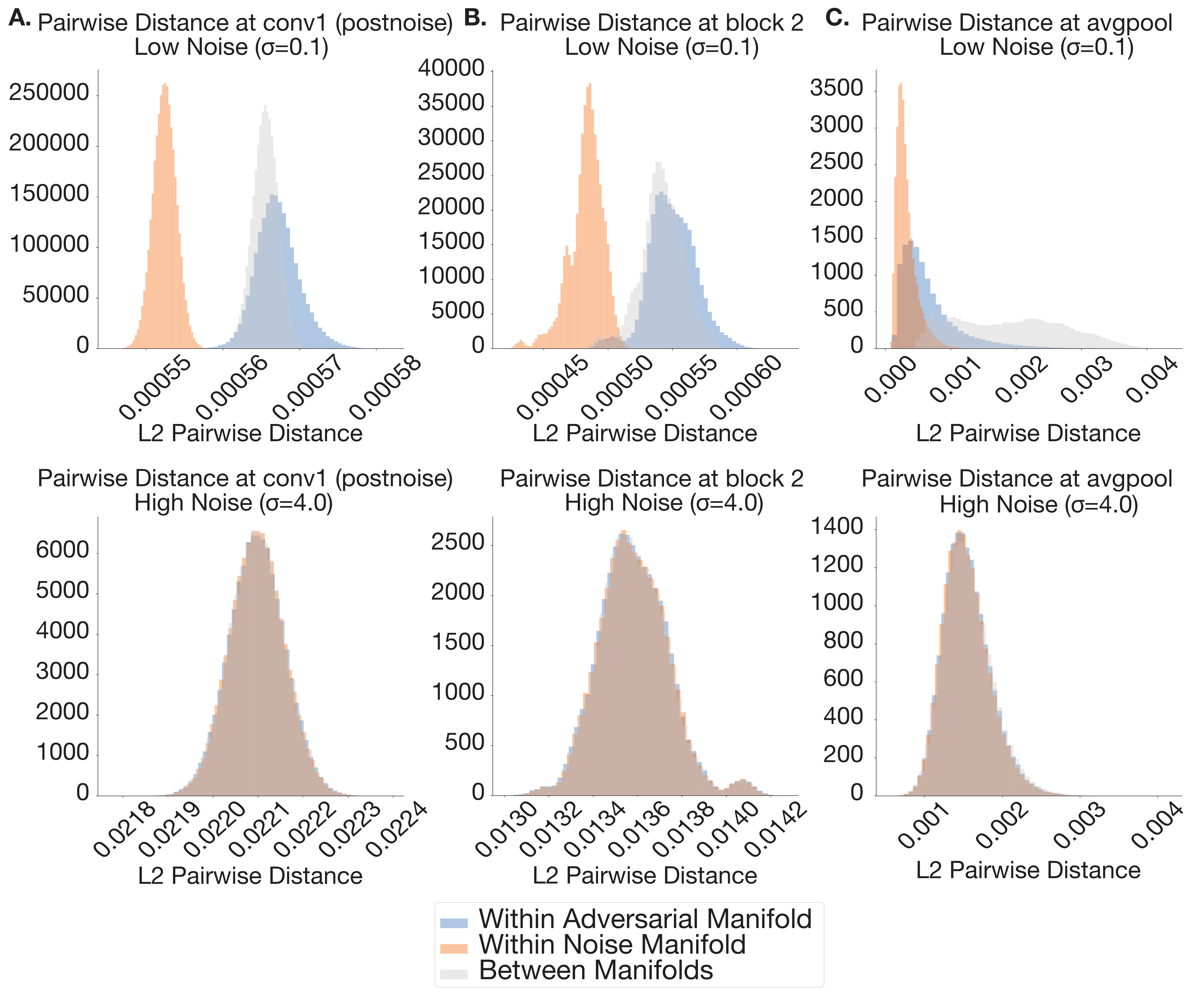}};
    \end{tikzpicture}
          \vspace{-4mm}
  \caption{\textbf{Gaussian noise model pairwise distance distributions} Pairwise distance distribution at layers \textbf{(A)} \texttt{conv1 (postnoise)}, \textbf{(B)} \texttt{block 2} and \textbf{(C)} \texttt{avgpool} of models with low noise level ($\sigma=0.1$) and high noise level ($\sigma=4.0$). At all measured layers, the adversarial exemplar and clean exemplar manifolds are more overlapped in the model with higher noise level ($\sigma=4.0$)}
  \label{fig:gaussian_pairwise}
\end{figure}

\begin{figure}[h!]
  \centering
    \begin{tikzpicture}
    \pgftext{\includegraphics[width=0.9\columnwidth]{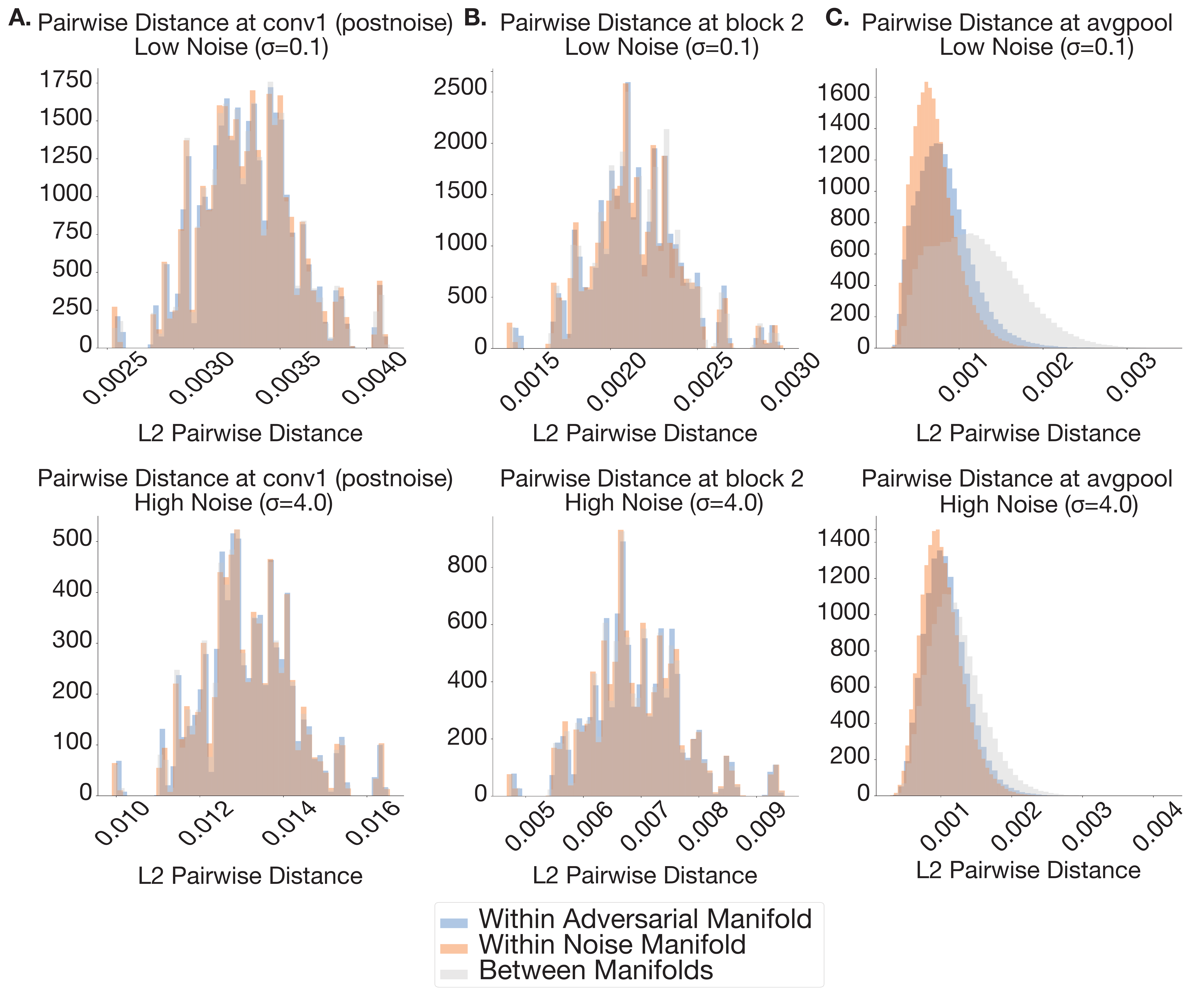}};
    \end{tikzpicture}
          \vspace{-4mm}
  \caption{\textbf{Pairwise distance distribution of Poisson-like noise model} Pairwise distance distribution at layers \textbf{(A)} \texttt{conv1 (postnoise)}, \textbf{(B)} \texttt{block 2} and \textbf{(C)} \texttt{avgpool}. At the early and intermediate layers (layer 2 and layer 4), the degree of manifold overlap is not visibly different between the low and high noise levels. However, at the late layer (layer 7), the model with high noise level has significantly more overlap between stochastic adversarial and clean manifolds than the model with lower levels of stochasticity. The level of stochasticity for the Poisson-like stochasticity model is defined by the variance/mean ratio. For standard Poisson, the variance/mean ratio is 1.}
  \label{fig:poisson_pairwise}
\end{figure}

This section provides an additional way of looking at the overlap between clean and adversarial exemplar manifolds using a pairwise distribution analysis of the Gaussian CIFAR-VOneNet model. To give a geometric intuition of the measure, Figure \ref{fig:pairwise_illustration} shows the relative positions of two manifolds in the feature space and its corresponding pairwise distance distribution in different conditions. In Figure \ref{fig:pairwise_illustration}A, the two manifolds are disjoint and both within-manifold distributions are not overlapped with the between-manifold distribution. In Figure \ref{fig:pairwise_illustration}B, the two manifolds are partly overlapped and the within-manifold distribution is also partly overlapped with the between-manifold distribution. Finally, in Figure \ref{fig:pairwise_illustration}C, when one of the manifolds is inclusive of the other, the mode of that within-manifold distribution will be larger than the mode of the between-manifold distribution. Figure \ref{fig:gaussian_pairwise} shows the pairwise distance distribution of three layers after the stochastic \texttt{conv 1 (postnoise)} layer, \texttt{block 1} layer and \texttt{avgpool} layer for low and high noise strength. A similar analysis is performed for the Poisson CIFAR-10 VOneNet in \ref{fig:poisson_pairwise}. 

\clearpage
\bibliographystyle{unsrt}
\bibliography{supplementary_material}

\clearpage